# Engineering in-plane anisotropy in 2D materials *via* surface-bound ligands

Tomoaki Sakurada,[1,2,3§] Woo Seok Lee,[1,4§] Yeongsu Cho,[1] Rattapon Khamlue,[5] Petcharaphorn Chatsiri,[5] Nicholas Samulewicz,[1] Tejas Deshpande,[1] Annlin Su,[1] Peter Müller,[6] Tadashi Kawamoto,[2] Shun Omagari,[2] Martin Vacha,[2] Watcharaphol Paritmongkol,[1,5] Heather J. Kulik,[1,6] William A. Tisdale[1]*

[1]Department of Chemical Engineering, Massachusetts Institute of Technology, Cambridge, Massachusetts 02139, United States

[2]Department of Materials Science and Engineering, Institute of Science Tokyo, Ookayama 2-12-1, Meguro-ku, Tokyo 152-8552, Japan

[3]Material Integration Laboratories, AGC Inc., Yokohama, Kanagawa 230-0045, Japan

[4]Department of Materials Science and Engineering, Massachusetts Institute of Technology, Cambridge, Massachusetts 02139, United States

[5]Department of Materials Science and Engineering, School of Molecular Science and Engineering, Vidyasirimedhi Institute of Science and Technology (VISTEC), Rayong 21210, Thailand

[6]Department of Chemistry, Massachusetts Institute of Technology, Cambridge, Massachusetts 02139, United States

[§]These authors contributed equally.

*Correspondence to: tisdale@mit.edu

**Abstract:**

2D materials exhibiting in-plane anisotropy enable novel functionality in electronic, optoelectronic, and photonic devices, yet their availability is generally limited to naturally-occurring low-symmetry van der Waals compounds. Here, we demonstrate an approach to structural engineering in a family of blue-emitting 2D silver phenylchalcogenide semiconductors based on steric interactions among surface-bound organic molecular ligands. By strategically halogenating specific sites of phenyl ligands, we demonstrate dramatic changes to the inorganic AgSe plane in mithrene (silver phenylselenolate, AgSePh). Density functional theory revealed pronounced in-plane electronic anisotropy for direct-gap fluorinated derivatives, while a chlorinated variant exhibited a direct-to-indirect bandgap transition. Furthermore, some fluorinated variants displayed strongly polarized absorption and luminescence, accompanied by a 10× enhancement in photoluminescence quantum yield. This work establishes a versatile approach for tailoring optoelectronic properties in hybrid semiconductors that is difficult or impossible to achieve in all-inorganic materials alone, offering new opportunities in advanced material design.

## Introduction

2D materials exhibiting in-plane anisotropy – defined by directional-dependent optical, electronic, magnetic, or vibrational properties – enable novel functionality in electronic, optoelectronic, and photonic devices such as waveplates, polarization-sensitive photodetectors, thermoelectrics, logic circuit elements, and advanced imaging technologies.[1–3] Despite the extensive exploration of widely studied 2D van der Waals materials such as graphene, boron nitride, and transition metal dichalcogenides (TMDs), their inherent in-plane isotropy, dictated by high-symmetry hexagonal lattices, imposes fundamental limitations for anisotropic applications.[4,5] Recent efforts have focused on low-symmetry 2D materials such as black phosphorus, SnSe, GaTe, and $ReSe_2$, which naturally break in-plane symmetry and offer promising pathways for achieving optical anisotropy.[6] Despite these advances, the realization of strong and tunable in-plane anisotropy remains a challenge, as inorganic crystals tend to stabilize in high-symmetry configurations. This limitation highlights the need for innovative material platforms that enable structural manipulation beyond conventional symmetry constraints. Moreover, the ability to decouple the degree of structural anisotropy from elemental composition of the inorganic layer would offer an added degree of engineering flexibility.

Hybrid organic-inorganic 2D materials, such as lead halide perovskites (LHPs) and metal organochalcogenides (MOCs), offer new opportunities to overcome these limitations by enabling structural and optoelectronic property tuning through organic ligand modification.[7–19] In these materials, steric interactions among organic ligands can induce strain in the underlying inorganic lattice, forcing inorganic elements into bonding configurations that are otherwise inaccessible. For instance, organic interlayer cations in 2D LHPs enable structural diversity not only in the stacking of $[PbX_6]^{2-}$ layers[8] but also in the configuration of $[PbX_6]^{2-}$ octahedra.[20] However, due to the weak interaction between the organic and inorganic components, the structural change in the inorganic framework induced by organic functionalization in 2D LHPs are typically limited to slight tilting or distortion of octaheral network, resulting in only marginal changes in electronic and optical properties.[21]

Metal organochalcogenides (MOCs) are another promising platform. Unlike other 2D van der Waals semiconductors and 2D perovskites, MOCs feature strong covalent bonding between the inorganic layer and organic ligands. Among them, 2D silver phenylselenide (**AgSePh**, or “mithrene”) (**Fig. 1a**) has attracted particular interest because of its narrow blue luminescence,[22] in-plane optical anisotropy,[23–25] ultrastrong light matter coupling,[26,27] p-type transport[28] and capability for optoelectronic property tuning through organic ligand modification.[29–32] Unmodified, **AgSePh** already exhibits some degree of in-plane optical anisotropy (**Fig. 1b**).[23–25] For example, Maserati *et al.* demonstrated that **AgSePh** exhibits three excitonic absorption states and emission is polarized emission along the [010] direction.[23] Jariwala and co-workers quantified its anisotropy using absorption linear dichroism, reporting values of 24.4%, which increased to 77.1% through cavity coupling.[24] These studies highlight the native optical anisotropy of **AgSePh**, and inspire efforts to further enhance in-plane anisotropy through ligand engineering.

In this study, we demonstrate the enhancement of optical anisotropy in mithrene through in-plane structural engineering of the AgSe layer, achieved via multi-site halogenated ligand substitution. We synthesized six mithrene variants with multi-site fluorinated (**$F_2$(2,3), $F_2$(2,4), $F_2$(2,5), $F_3$(2,3,4) $F_3$(2,4,5)**, **Fig. 1c**) and chlorinated ligands (**$Cl_2$(2,3)**, **Fig. 1d**). Single-crystal X-ray diffraction reveal remarkable structural changes in the AgSe plane between **AgSePh**, **$F_2$(2,3)/$F_2$(2,4)/$F_2$(2,5)** and **$Cl_2$(2,3)** (**Fig. 1e**), driven by steric hinderance and inter molecular interaction between neighboring ligands. Density functional theory calculations reveal significantly pronounced in-plane anisotropic electronic band sturcture in **$F_2$(2,3)/$F_2$(2,4)/$F_2$(2,5)** compared to AgSePh and indirect band structure in **$Cl_2$(2,3)**. Steady-state and time-resolved photoluminescence spectroscopy show narrow blue emission with an order-of-magnitude improvement in photoluminescence quantum efficiency (PLQY) of **$F_2$(2,3)** and **$F_2$(2,5)** compared to AgSePh, while **$Cl_2$(2,3)** shows no measurable light emission. Furthermore, **$F_2$(2,3)** (**Fig. 1f**) exhibits strongly polarized absorption and luminescence properties along the [100] direction (**Fig. 1g**), with absorption linear dichroism exceeding 80%. This study establishes a versatile strategy for precise in-plane structural tuning of 2D hybrid semiconductors, paving the way for the development of materials with enhanced optoelectronic properties and advanced anisotropic functionalities.

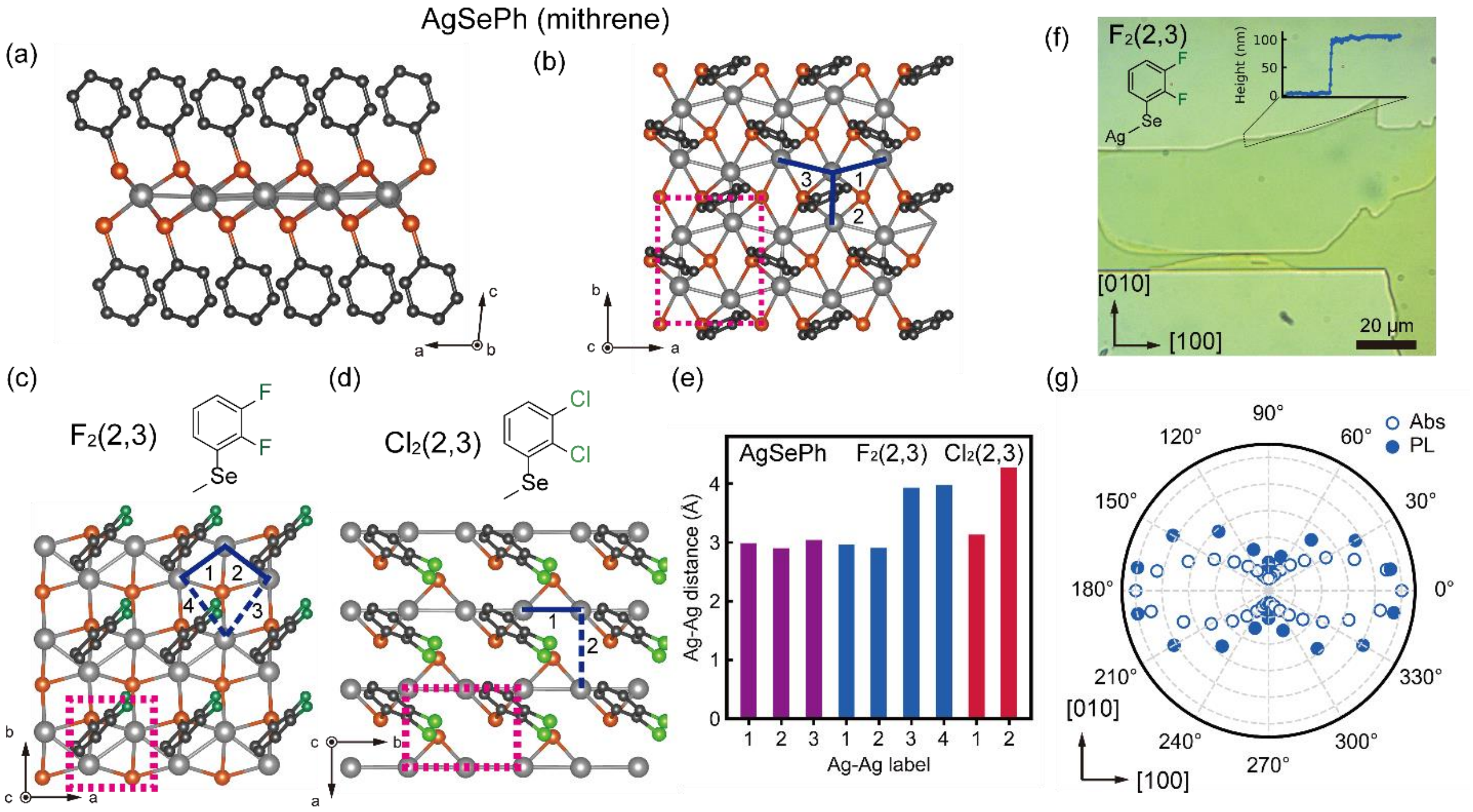

**Fig 1. Structure of 2D MOCs.** (a) Side view of **AgSePh**. Top view for (b) AgSePh, (c) **$F_2$(2,3)**, and (d) **$Cl_2$(2,3)**. Ag, Se, C atoms are depicted in gray, orange, and black, respectively. Dotted square represents the primitive cell, and blue lines represent nearest Ag-Ag distances. (e) Distance between Ag atoms for each label. (f) Optical micrograph of exfoliated **$F_2$(2,3)** single crystal. (g) Polar plot of absorption (hollow circles) and photoluminescence (PL, filled circles) intensity of **$F_2$(2,3)** single crystal at peak resonance.

## Results and Discussion

**Synthesis and Characterization of 2D MOCs.**

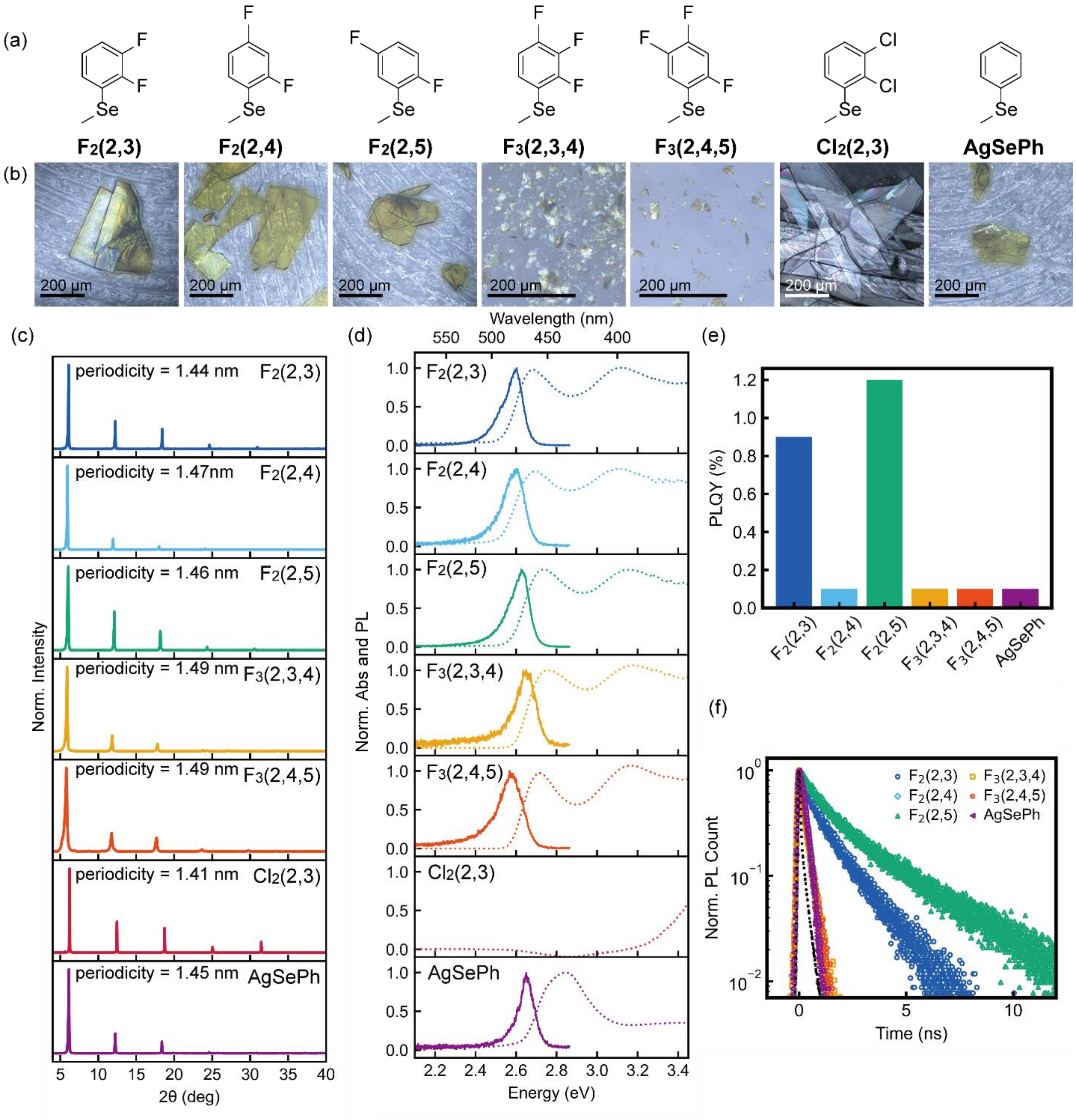

**Fig 2. Diffraction and Optical Properties of Synthesized MOCs.** (a) Molecular structures of ligands. (b) Optical microscope images of corresponding MOC crystals. (c) PXRD pattern of MOC crystals, showing characteristic interlayer diffraction peaks. (d) Normalized absorption (dotted) and PL (solid) spectra of each compound. (e) Absolute PLQY of each compound at room temperature. (f) Time-resolved PL decay of each compound (room temperature). Black dotted line corresponds to instrument response function.

Recently, our group reported synthesis and characterization of *mono*-substituted mithrene variants, demonstrating precise optical band gap tuning without largely changing the Ag-Se layer structure.[33] Here, we examined the effect of multi-site substitution by halogen atoms, including at least one *ortho*-position. **Fig. 2a** showcases the halogenated benzeneselenol ligands investigated, along with the parent **AgSePh** as a reference. These novel ligands include three difluoro-benzeneselenols, two trifluoro-benzeneselenols, and one dichloro-benzeneselenol, which we refer to as **$F_2$(2,3)**, **$F_2$(2,4)**, **$F_2$(2,5)**, **$F_3$(2,3,4)**, **$F_3$(2,4,5)**, and **$Cl_2$(2,3)**, respectively.

Halogenated **AgSePh** crystals were prepared using a modified amine-assisted crystallization method.[32] Silver nitrate ($AgNO_3$) and halogenated organodiselenides – synthesized *via* Grignard reactions (See Supporting Information Section) – were dissolved in 1-butylamine. The solutions were combined in a vial and the mixture was placed inside a jar containing deionized water, which works as antisolvent, leading to crystal formation in ~6 days. Submillimeter lateral sized crystals were obtained for **$F_2$(2,3)**, **$F_2$(2,4)**, **$F_2$(2,5)**, **$Cl_2$(2,3)** and **AgSePh**, whereas **$F_3$(2,3,4)** and **$F_3$(2,4,5)** formed aggregates of microcrystals (**Fig. 2b**).

Powder X-ray diffraction (PXRD) analysis of ground crystals suggests that all mithrene derivatives exhibit 2D van der Waals crystals with an interlayer separation of 1.4-1.5 nm, similar to **AgSePh** (**Fig. 2c**). Thermogravimetric analysis revealed that they are thermally stable up to at least 220 °C under $N_2$ atmosphere (**Fig. S2**). Residual weight following thermal decomposition corresponds to the calculated Ag contents, further supporting the purity of synthesized materials.[34,35] The infrared spectra of mithrene variants reveal differences in the 1000-1500 $cm^{-1}$ region, which can be attributed to the symmetric and asymmetric stretching modes of C–H or C–F bonds (**Fig. S3**).[36]

**Fig. 2d** displays absorption and photoluminescence (PL) spectra of multi-site halogenated compounds and **AgSePh** at room temperature. Fluorinated mithrenes exhibit similar absorption spectra with two distinct peaks having large energy separation, centered at ~2.7 eV and 3.1 eV. In contrast, **$Cl_2$(2,3)** exhibits significantly blue-shifted absorption spectra without pronounced peaks. Notably, the absorption features of halogenated mithrenes are distinct from those of **AgSePh**, where multiple excitonic resonances are crowded in the range between 2.6 eV and 2.9 eV.

Fluorinated mithrene derivatives exhibit narrow blue emission at room temperature upon 405 nm photoexcitation, with emission peaks centered at 2.57-2.65 eV and full-width at half-maximum (FWHM) of 108 to 150 meV, similar to **AgSePh** which shows luminescence centered at 2.67 eV with FWHM of 91 meV (**Fig. 2d** and **Table S2**). In contrast, **$Cl_2$(2,3)** did not show measurable luminescence upon 375 nm photoexcitation at room temperature.

The PLQY of fluorinated mithrenes and **AgSePh** were measured using the absolute method in an integrating sphere at room temperature (**Fig. 2e**).[37] **$F_2$(2,3)** and **$F_2$(2,5)** exhibited PLQY of ~1%, representing an order-of-magnitude improvement over **AgSePh** (~0.1%), while **$F_2$(2,4)**, **$F_3$(2,3,4)** and **$F_3$(2,4,5)** showed lower PLQY of ~0.1% (**Fig. S4**). Moreover, single exponential fits to PL decay traces revealed increased PL lifetimes of **$F_2$(2,3)** and **$F_2$(2,5)** (0.5 ns and 0.7 ns, respectively) compared to **AgSePh**, **$F_2$(2,4)**, **$F_3$(2,3,4)** and **$F_3$(2,4,5)** that exhibited much shorter PL lifetimes of ~0.1 ns (**Fig. 2f**), suggesting a reduced non-radiative recombination rate in **$F_2$(2,3)** and **$F_2$(2,5)**. We note that PLQY of mithrene and its variants improves to >10% when cooled to cryogenic temperature, and the observed low efficiency of emission is consistent with nonradiative recombination arising from a residual background charge carrier density or defect sites, as observed in other 2D semiconductors.[14,18,38]

**Table 1.** Crystallographic Data for **$F_2$(2,3)**, **$F_2$(2,4)**, **$F_2$(2,5)**, **$Cl_2$(2,3)**, and **AgSePh**

| | **$F_2$(2,3)** | **$F_2$(2,4)** | **$F_2$(2,5)** | **$Cl_2$(2,3)** | **AgSePh**[32] |
|---|---|---|---|---|---|
| Empirical formula | $C_6H_3F_2AgSe$ | $C_6H_3F_2AgSe$ | $C_6H_3F_2AgSe$ | $C_6H_3Cl_2AgSe$ | $C_{12}H_{10}Ag_2Se_2$ |
| $M_r$ | 299.91 | 299.91 | 299.91 | 332.81 | 527.86 |
| Temperature (K) | 266 | 100(2) | 100(2) | 100(2) | 100(2) |
| Wavelength (Å) | 0.71069 | 0.71073 | 0.71073 | 0.71073 | 0.71073 |
| Crystal system | Monoclinic | Triclinic | Triclinic | Monoclinic | Monoclinic |
| Space group | $P2_1/n$ | $P$-1 | $P$-1 | $P2_1$ | $P2_1/c$ |
| $a$ (Å) | 4.7321(16) | 4.6960(2) | 4.7267(6) | 4.2844(3) | 5.8334(5) |
| $b$ (Å) | 4.8766(15) | 4.8270(3) | 4.8593(5) | 6.2929(5) | 7.2866(6) |
| $c$ (Å) | 28.866(8) | 29.3273(15) | 14.7295(17) | 14.1563(12) | 29.079(3) |
| $\alpha$ (°) | 90 | 90.3602(19) | 96.097(4) | 90 | 90 |
| $\beta$ (°) | 94.26(2) | 91.8486(18) | 96.655(4) | 93.505(3) | 95.5819(16) |
| $\gamma$ (°) | 90 | 91.9161(17) | 92.120(4) | 90 | 90 |
| $V$ (Å$^3$) | 664.3(4) | 664.04(6) | 333.72(7) | 380.96(5) | 1230.16(18) |
| $Z$ | 4 | 4 | 2 | 2 | 4 |
| Calculated density (Mg/m$^3$) | 2.999 | 3.000 | 2.985 | 2.901 | 2.850 |
| Absorption coefficient (mm$^{-1}$) | 8.449 | 8.460 | 8.417 | 8.031 | 9.067 |
| $F$(000) | 552 | 552 | 276 | 308 | 976 |
| Crystal size (mm$^3$) | 0.450 × 0.240 × 0.010 | 0.270 × 0.170 × 0.010 | 0.150 × 0.080 × 0.010 | 0.465 × 0.101 × 0.010 | 0.230 × 0.220 × 0.020 |
| $\theta$ range for data collection (°) | 2.830. to 27.588 | 2.084 to 31.521° | 1.401 to 32.137 | 2.883 to 30.058 | 2.815 to 31.540 |
| Index ranges | $-6 \leq h \leq 6$, $-6 \leq k \leq 6$, $-37 \leq l \leq 37$ | $-6 \leq h \leq 6$, $-7 \leq k \leq 7$, $-43 \leq l \leq 43$ | $-7 \leq h \leq 7$, $-6 \leq k \leq 7$, $-21 \leq l \leq 21$ | $-6 \leq h \leq 5$, $-8 \leq k \leq 8$, $-19 \leq l \leq 19$ | $-8 \leq h \leq 8$, $-10 \leq k \leq 10$, $-42 \leq l \leq 42$ |
| Reflections collected | 2163 | 45018 | 18147 | 10659 | 37922 |
| Independent reflections | 1618 [$R_{int}$ = 0.0718] | 4382 [$R_{int}$ = 0.0388] | 2319 [$R_{int}$ = 0.0450] | 2172 [$R_{int}$ = 0.0412] | 4107 [$R_{int}$ = 0.0381] |
| Completeness to $\theta$ = 25.242° | 99.2% | 99.2% | 99.5 % | 99.8% | 99.5 % |
| Data / restraints / parameters | 1618 / 18 / 80 | 4382 / 1052 / 327 | 2319 /420/152 | 2172 / 1 / 93 | 4107 / 32 / 145 |
| Goodness-of-fit on $F^2$ | 1.100 | 1.279 | 1.293 | 1.088 | 1.155 |
| Final R indices [$I > 2\sigma(I)$] | $R1$ = 0.0885, $wR2$ = 0.2324 | $R1$ = 0.0518, $wR2$ = 0.1236 | $R1$ = 0.0588, $wR2$ = 0.1488 | $R1$ = 0.0162, $wR2$ = 0.0356 | $R1$ = 0.0341, $wR2$ = 0.1051 |
| R indices (all data) | $R1$ = 0.1683, $wR2$ = 0.2892 | $R1$ = 0.0555, $wR2$ = 0.1248 | $R1$ = 0.0645, $wR2$ = 0.1512 | $R1$ = 0.0208, $wR2$ = 0.0368 | $R1$ = 0.0393, $wR2$ = 0.1090 |
| Largest diff. peak and hole (e.Å$^{-3}$) | 3.80 and −2.19 | 1.567 and −2.063 | 3.137 and −2.459 | 0.746 and −0.793 | 1.817 and −2.419 |

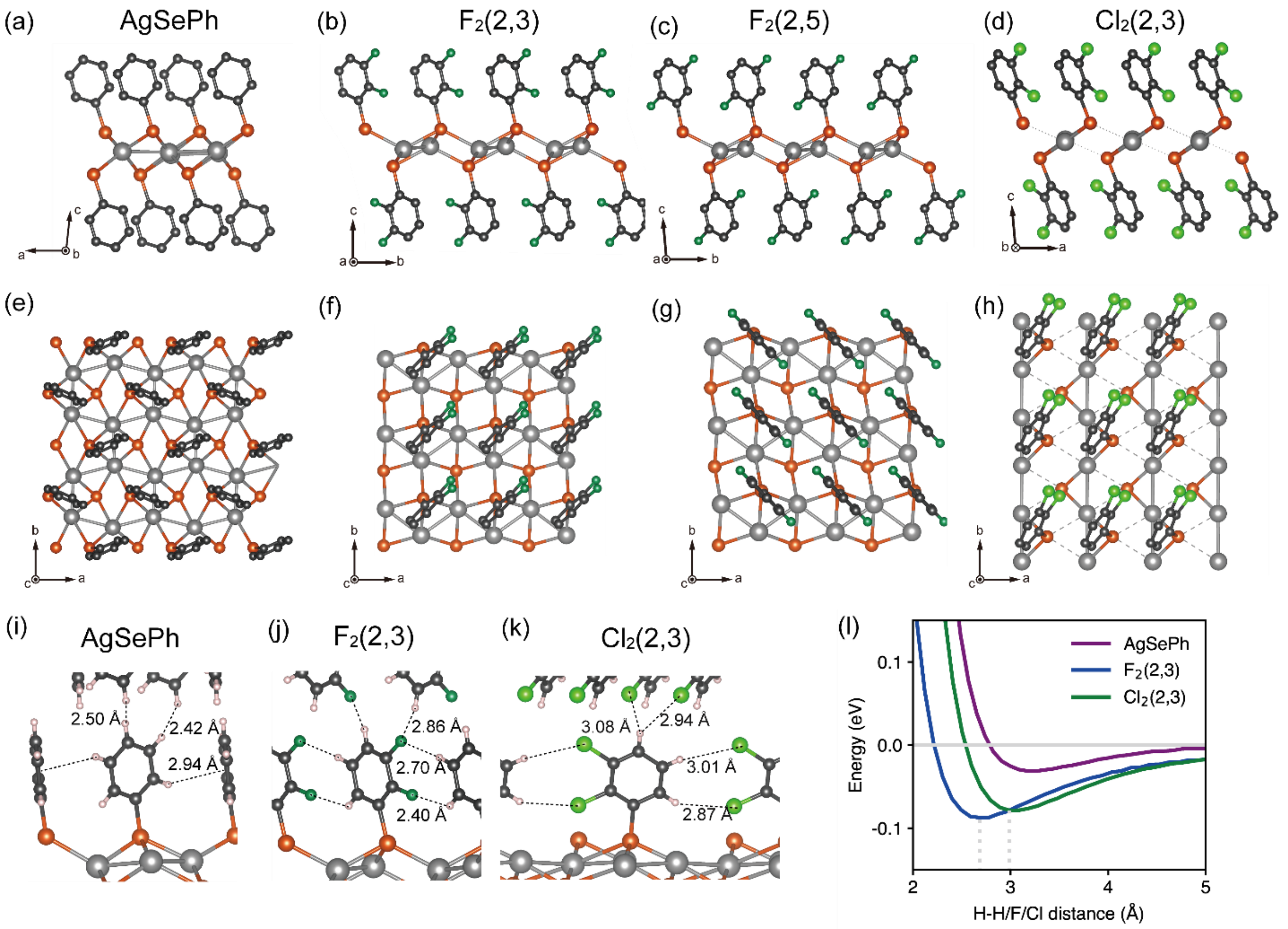

**Figure 3. Single-crystal structures of 2D halogenated silver phenylselenides.** Layered 2D structure from side and top view for **AgSePh** (a,e), **F2(2,3)** (b,f), **F2(2,3)** (c,g) and **Cl2(2,3)** (d,h). Ag, Se, C, F, and Cl atoms are depicted by gray, orange, black, green, and light green spheres, respectively. Hydrogen and distorted atoms are omitted for clarity. Distance between adjacent ligand for (i) **AgSePh**, (j) **F2(2,3)** and (k) **Cl2(2,3)**. (l) Interaction energy between two ligands as a function of the H-H (purple), H-F (blue), or H-Cl (green) distance, calculated using the PBE functional with a plane-wave basis set. The dashed gray vertical line denotes the distance corresponding to the energy minimum, which are 2.68 Å for **F2(2,3)** and 2.99 Å for **Cl2(2,3)**, aligning well with the H-F and H-Cl separations observed within the MOC.

Single crystal XRD analysis revealed distinct structural features across the investigated mithrene variants. **$F_2$(2,3)** and **$Cl_2$(2,3)** crystallize in monoclinic centrosymmetric space groups $P2_1/n$ and non-centrosymmetric $P2_1$, respectively, while **$F_2$(2,4)** and **$F_2$(2,5)** adopt triclinic $P1$ symmetry (**Table 1**). Representative crystal structures of **AgSePh**, **$F_2$(2,3)**, **$F_2$(2,5)**, and **$Cl_2$(2,3)** are illustrated in **Fig. 3**, highlighting their characteristic 2D van der Waals architectures composed of Ag-Se layers surrounded with organoselenol ligands (**Figs. 3a–d**). Unlike *mono*-substitution of **AgSePh**,[39] distinct variations in the Ag-Se layer geometry and ligand arrangements are observed (**Figs. 3e–h** and **S5**).

For **AgSePh**, the Ag-Ag interactions within the Ag-Se layers are uniformly distributed along both the *a*- and *b*-axes, with an interatomic distance of ~3 Å (**Figs. 3e** and **S5**). Se atoms coordinate to four Ag atoms with Ag-Se bond lengths ranging from ~2.6–2.7 Å (**Fig. S5h**). In contrast, the introduction of fluorine atoms in the ligands induces structural expansion in the Ag-Se layers, resulting in a rectangular grid-like pattern of Ag and Se atoms (**Figs. 3f, 3g**, and **S5**). Here, Ag-Ag interactions are restricted to the *a*-axis, forming a zigzag motif, while the *b*-axis exhibits extended Ag-Ag distances (~4 Å; **Fig. S5g**), disrupting the Ag-Ag network along this direction. Structural analogies among fluorinated mithrenes are observed, albeit with slight variations in Ag-Ag and Ag-Se bond lengths (**Fig. S6**).

In **$Cl_2$(2,3)**, Ag atoms align linearly along the *b*-axis with interatomic distances of 3.1 Å (**Fig. 3h**). Unlike **AgSePh** and fluorinated mithrenes, Se atoms in **$Cl_2$(2,3)** coordinate to only two Ag atoms, forming zigzag Ag-Se chains that stack to generate layered structures (**Figs. S5** and **S7**). The interchain Ag-Se distance (~3.2 Å) suggests weak or negligible covalent interactions between adjacent chains.

Ligand packing further distinguishes the structural variants. **AgSePh** exhibits a herringbone arrangement of phenyl rings (**Fig. 3e**), whereas halogenated mithrenes display a brickwork alignment of benzene rings (**Figs. 3f–h**). This parallel orientation is likely stabilized by hydrogen bonding interactions between halogen atoms (F or Cl) and neighboring hydrogen atoms, as indicated by short F-H or Cl-H distances (~2.4–3.0 Å; **Figs. 3i–k**). Computational studies support the energetic favorability of linear ligand arrangements in **$F_2$(2,3)** and **$Cl_2$(2,3)**, consistent with the crystallographic observations (**Fig. 3l**).

## Electronic structure calculations

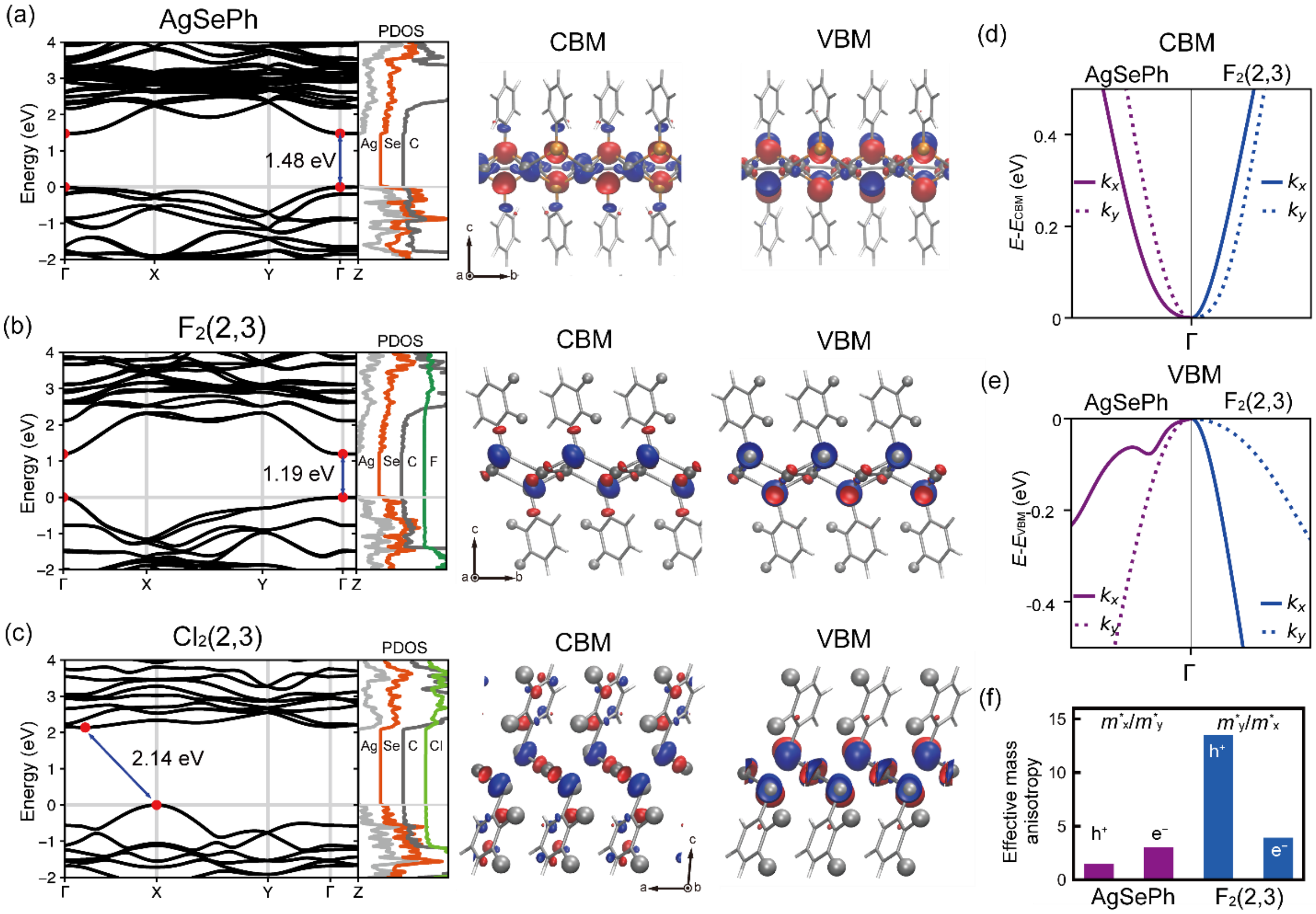

**Figure 4. Electronic Structure Calculations.** Calculated electronic band structure, projected density of states (PDOS), and wavefunctions of the conduction band minimum (CBM) and valence band maximum (VBM) for (a) **AgSePh**, (b) **F2(2,3)** and (c) **Cl2(2,3)**. Red dots mark the locations of the VBM and CBM. C and H atoms are represented with a stick model, and Ag, Se, F, and Cl are shown as spheres. The red and blue surfaces represent the positive and negative signs of the wavefunctions at an isosurface level of 0.003 $Å^{-3}$. Comparison of the band edge of (d) CBM and (e) VBM between **AgSePh** and **F2(2,3)** along the in-plane *x* ([100]) and *y* ([010]) direction. (f) Ratio of effective masses along the in-plane *x* ([100]) and *y* ([010]) directions.

To elucidate the impact of structural differences in the Ag-Se layer of mithrene variants on their electronic band structures, we performed plane-wave, semi-local DFT calculations (**Fig. 4**). Fluorinated MOCs exhibit band structures that closely resemble **AgSePh**, showing direct band gaps in the range of 1.16-1.22 eV, which are smaller than the 1.48 eV band gap of **AgSePh** (**Figs. 4a, 4b, and S8**). In contrast, **$Cl_2$(2,3)** shows an indirect and widened band gap of 2.14 eV, consistent with its lack of emission under UV excitation and blue-shifted absorption spectra (**Fig. 4c**). The difference can be attributed to the coordination environment of Ag atoms in **$Cl_2$(2,3)**, where each Ag atom binds to two Se atoms instead of fourfold coordination observed in **AgSePh** and **$F_2$(2,3)**. The reduced coordination weakens orbital hybridization, resulting in a less dispersive conduction band and a larger bandgap in **$Cl_2$(2,3)**. All MOCs have negligible band dispersion along the out-of-plane direction (Γ→Z), confirming strong confinement of charge carriers within the 2D plane.

The frontier orbitals at the conduction band minimum (CBM) and valence band maximum (VBM) are depicted alongside the electronic band structures (**Figs. 4**, **S9** and **10**). **AgSePh**, **$F_2$(2,3)**, **$F_2$(2,4)**, and **$F_2$(2,5)** all exhibit conduction and valence band edges dominated by Ag and Se orbitals. In contrast, for **$Cl_2$(2,3)**, the CBM exhibits a significant contribution from the organic ligands due to its enlarged band gap.

The band dispersion of **AgSePh** and **$F_2$(2,3)** along the two principal crystal orientations are shown in **Fig. 4d** and **4e**. The conduction and valence band of **AgSePh** show strong dispersion along both *x* and *y* axis, with the ratio of effective masses for holes and electrons ($m_x^*/m_y^*$) being 1.3 and 3.0, respectively (**Fig. 4f**, **Table S3**). In comparison, **$F_2$(2,3)** demonstrates significant valence band (VB) dispersion along the *x*-axis ([100] direction) but dramatically reduced band dispersion along the *y*-axis ([010] direction), with the ratio of effective masses for holes ($m_y^*/m_x^*$) reaching 13.5 and for electrons, 3.9. Reduced band dispersion along the *y*-axis is a direct consequence of the strain imparted to the Ag-Se lattice by steric interactions between the halogenated phenyl ligands.

## Optical anisotropy of 2D MOCs

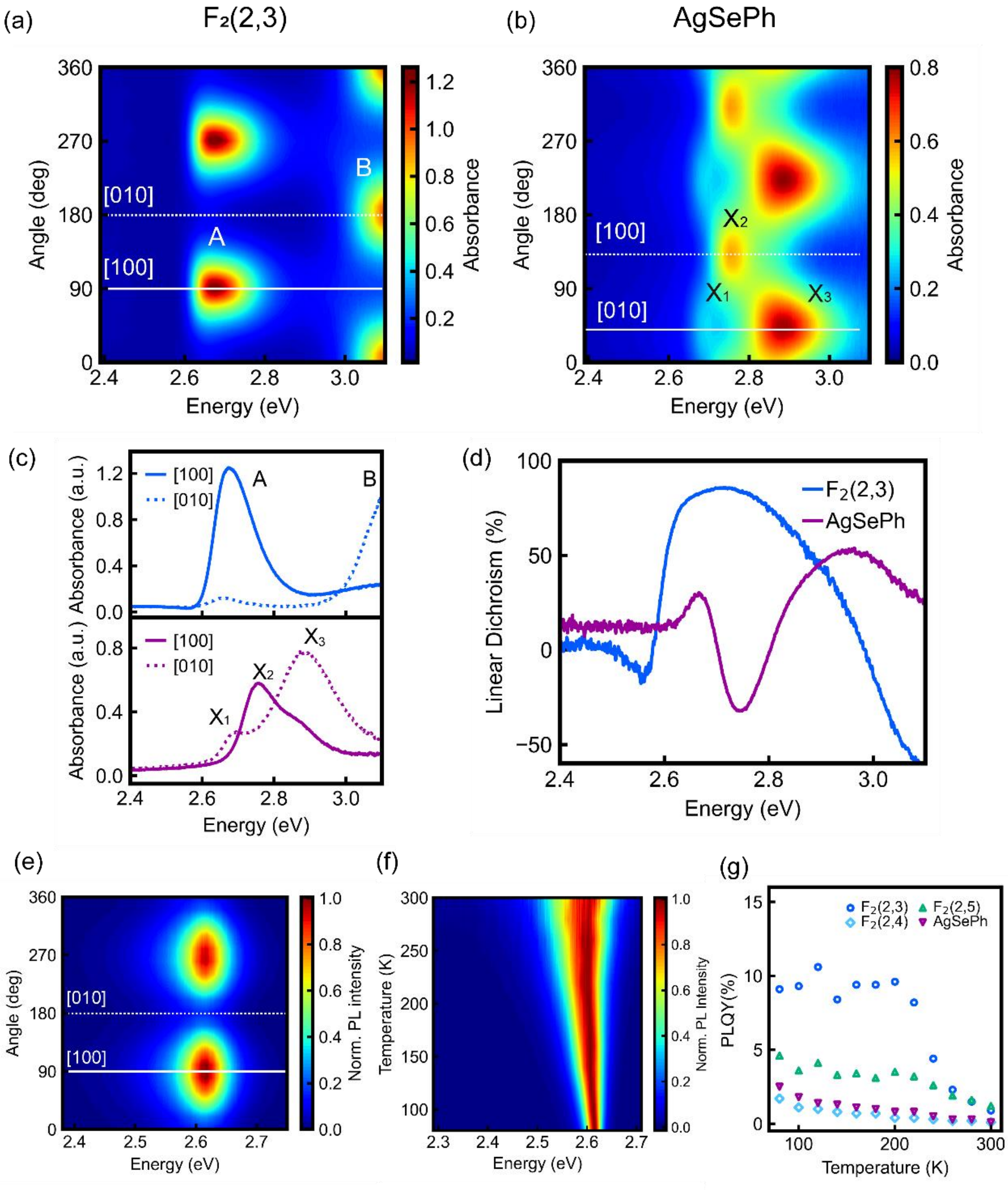

**Figure 5. Optical Anisotropy of $F_2$(2,3).** Absorption spectra of (a) **$F_2$(2,3)** and (b) **AgSePh** as a function of polarization angle of incident light. (c) Absorption spectra of **$F_2$(2,3)** (blue) and **AgSePh** (purple) with polarization parallel to the [100] or [010] direction. (d) Linear dichroism of **$F_2$(2,3)** and **AgSePh** as a function of photon energy. (e) PL spectra of **$F_2$(2,3)** as a function of polarization angle of the emitted light. (f) Temperacture dependent PL spectra of **$F_2$(2,3)** from 80 to 300 K. (g) Temperature deoendent PLQY for $F_2$ MOCs and **AgSePh**.

In addition to its effect on electronic band dispersion, ligand-induced strain also dramatically affects the anisotropy of optical absorption in halogenated mithrene variants. Absorption anisotropy at room temperature was studied using polarization-resolved micro-absorption spectroscopy in transmission mode with incident light polarization controlled by a linear polarizer (See Methods for details). A 100 nm-thick single-crystalline **$F_2$(2,3)** flake was prepared through mechanical exfoliation to avoid additional complications from photonic effects that may arise in thicker crystals (**Fig. 1f**). The single crystallinity of the flake was confirmed by cross-polarized optical microscopy (**Fig. S11**). The absorption and PL anisotropy were correlated with crystal orientation using transmission electron microscopy and electron diffraction measurements (**Fig. S12**).

**Fig. 5a** shows the linear-polarization-resolved absorption spectra of **$F_2$(2,3)**, revealing strong absorption anisotropy distinct from **AgSePh** (**Fig. 5b**). When the incident light polarization was aligned with the crystallographic [100] direction, the absorption peak at ~2.67 eV (A) was maximized while the absorption peak at ~3.10 eV (B) was significantly suppressed (**Fig. 5c**). Conversely, with the light polarization along the [010] direction, the absorption intensity of feature A was minimized while the intensity of feature B became dominant, indicating that the two features are orthogonally polarized (A||[100] and B||[010], **Fig. 5a** and **5c**). In contrast, **AgSePh** exhibited three closely-spaced absorption peaks at ~2.69 eV ($X_1$), ~2.76 eV ($X_2$), and ~2.89 eV ($X_3$), each showing some degree of in-plane anisotropy ($X_1$/$X_3$||[010] and $X_2$||[100], **Fig. 5b** and **5c**) . (NOTE: $X_1$ feature in AgSePh is further resolved to $X_{1a}$(||[100]) and $X_{1b}$||[010]) at cryogenic temperature)[25]

To compare the absorption anisotropy between **AgSePh** and **$F_2$(2,3)**, we calculated the linear dichroism (LD) as a function of photon energy, defined as,

$$LD = \frac{A_{[100]} - A_{[010]}}{A_{[100]} + A_{[010]}}$$

where $A_{[100]}$ and $A_{[010]}$ correspond to the absorption intensity when the light polarization was aligned along the crystallographic [100] and [010] directions, respectively (**Fig. 5d**). **$F_2$(2,3)** exhibited significantly stronger LD – 85% at ~2.71 eV (457 nm) and 63% at ~3.10 eV (400 nm) – over a broad wavelength range (|LD|>40 for 426-477 nm and 400-408nm; note that 400 nm is cut-off wavelength of our measurement) – in contrast to **AgSePh**, which showed 30% at ~2.66 eV, 33% at ~2.75 eV, and 55% at ~2.95 eV (**Fig. 5d**). The enhanced linear dichroism in **$F_2$(2,3)** arises from the the combination of a large energetic separation of absorption peaks and strong in-plane anisotropy of each optical transition. Notably, most in-plane LD values reported for 2D materials to date fall below 40%, and LD exceeding 60% in the near-UV to visible range is particularly rare,[1,5,40] underscoring the signficance of the optical anisotropy enhancement demonstrated in this work.

**Fig. 5e** displays linear-polarization-resolved emission spectra of **$F_2$(2,3)**, measured by rotating a linear polarizer in the collection path upon circularly polarized 405 nm photo-excitation. **$F_2$(2,3)** exhibited strong emission anisotropy polarized along the [100], consistent with the lowest absorption peak (feature A), with a slight red-shift of ~55 meV. The PL intensity at the peak position was fitted well with a curve using the formula of $A = (A_{\max} - A_{\min})\cos^2(\theta - \theta_0) + A_{\min}$, where $\theta_0$ denotes the reference polarization angle (**Fig. S13**). The dichroism ratio of PL intensity, defined as $A_{\max}/A_{\min}$, is ~6 for **$F_2$(2,3)** and ~4 for **AgSePh**, indicating strong PL anisotropy of **$F_2$(2,3)**. We note that the absorption and photoluminescence (PL) polarization in **$F_2$(2,3)** are consistent with strong valence and conduction band dispersion in the [100] direction, which also corresponds to the direction of strong nearest-neighbor coupling in **$F_2$(2,3)** (**Figs. 5a, 5e**).

Temperature-dependent photoluminescence spectroscopy of fluorinated mithrene variants was performed over the temperature range of 80 to 300 K (**Figs. 5f** and **S14**). Upon lowering the temperature, the PL peak blue-shifted and narrowed for all materials investigated (**Fig. S14**). We estimated the PLQY at lower temperature by scaling the absolute PLQY at 300 K by relative intensity at reduced temperature (**Fig. 5g**). The PLQY of **$F_2$(2,3)** and **$F_2$(2,5)** increased with decreasing the temperature, reaching near saturation at ~200 K at 10% and 5%, respectively. Temperature-dependent transient PL exibited non- or multi-exponential decay dynamics, suggesting exciton scattering among multiple near-band-edge states in these systems (**Fig. S15**).

## Conclusion

In summary, we demonstrated a successful strategy for engineering the in-plane structure in a blue-emitting 2D van der Waals semiconductor through halogenation of pendant organic ligands. Specifically, by halogenating the *ortho*- and *meta*- positions of phenyl ligands in **AgSePh**, we achieved enhanced optical anisotropy in fluorinated mithrene **$F_2$(2,3)** and band gap modulation in **$Cl_2$(2,3)**. Structural and theoretical analyses revealed that the incorporation of multiple halogen atoms induces changes in the Ag-Se layers driven by steric hindrance and hydrogen bonding interactions among phenyl ligands. Fluorinated mithrene showed blue luminescence, with **$F_2$(2,3)** and **$F_2$(2,5)** achieving 10-fold enhancement in PLQY compared with **AgSePh**, whereas **$Cl_2$(2,3)** displayed no emission at room temperature. Further optical characterization revealed that both absorption and emission from **$F_2$(2,3)** are strongly polarized along the [100] direction, with linear dichroism exceeding 80% in absorption. This work establishes a versatile structural engineering strategy for tailoring the optoelectronic properties of hybrid semiconductor systems that is difficult or impossible to achieve in all-inorganic materials alone, offering new opportunities in advanced material design.

## ASSOCIATED CONTENT

### Supporting Information

The Supporting Information is available

> Experimental details, synthesis and characterization of organic precursors, additional structural and characterization data for synthesized MOCs, and theoretical calculation results (PDF).

### Accession Codes

Deposition Numbers 2501571 (**$F_2$(2,3)**), 2502498 (**$F_2$(2,4)**), 2502499 (**$F_2$(2,5)**), 2412944 (**$Cl_2$(2,3)**), 2501947 (**1**), 2501948 (**2**), 2501952 (**4**), 2501953 (**5**) and 2473793 (**6**) contain the supplementary crystallographic data for this paper. These data can be obtained free of charge via www.ccdc.cam.ac.uk/data_request/cif, or by emailing data_request@ccdc.cam.ac.uk, or by contacting The Cambridge Crystallographic Data Centre, 12 Union Road, Cambridge CB2 1EZ, UK; fax: +44 1223 336033

## AUTHOR INFORMATION

### Corresponding Author

*E-mail: tisdale@mit.edu

### ORCID

| | |
|---|---|
| Tomoaki Sakurada | 0000-0002-2353-7324 |
| Woo Seok Lee | 0000-0001-9188-5104 |
| Yeongsu Cho | 0000-0001-8159-600X |
| Rattapon Khamlue | 0009-0007-1733-8437 |
| Petcharaphorn Chatsiri | 0009-0008-4514-5314 |
| Nicholas Samulewicz | 0009-0003-2981-442X |
| Tejas Deshpande | |
| Annlin Su | 0000-0002-4287-9047 |
| Peter Müller | 0000-0001-6530-3852 |
| Tadashi Kawamoto | 0000-0002-5676-4013 |
| Shun Omagari | 0000-0002-7128-0295 |
| Martin Vacha | 0000-0002-5729-9774 |
| Watcharaphol Paritmongkol | 0000-0003-1638-6828 |
| Heather J. Kulik | 0000-0001-9342-0191 |
| William A. Tisdale | 0000-0002-6615-5342 |

**Data availability**

All available data are included in the main article and ESI.

**Author contributions**

All authors discussed the results and reviewed the manuscript.

**Conflicts of interest**

There are no conflicts to declare.

**Acknowledgements**

Synthesis and structural characterization of silver phenylchalcogenides at MIT was supported by the U.S. Army Research Office under Award Number W911NF-23-1-0229 (T.S., W.S.L., N.S., T.D., W.P., and W.A.T.). Crystal growth, structural characterization, and diffuse reflectance spectroscopy at VISTEC were supported by the Development of International Cooperation Network on Science Technology and Innovation, Program Management Unit for Human Resources & Institutional Development, Research and Innovation (PMU-B) of Thailand under the grant number B42G680032. Spectroscopic characterization at MIT was partially supported by the U.S. Department of Energy, Office of Science, Basic Energy Sciences under award number DE-SC0019345 (T.S., W.S.L.,W.A.T.). Density functional theory calculations at MIT were supported by the U.S. Department of Energy under grant number DE-SC0024174 (Y.C. and H.J.K.). The authors acknowledge the MIT SuperCloud for providing HPC resources that have contributed to the research results reported within this paper. M.V. was supported by the SPS KAKENHI grant number 23H04875 in Grant-in-Aid for Transformative Research Areas 'Materials Science of Meso-Hierarchy'. T.S. acknowledges the financial support from AGC Inc. and is also grateful to the supervisors at AGC, Mr. Shin Tatematsu and Dr. Yoshitomi Morizawa for their initial support and continuous encouragement. R.K. and P.C. acknowledge the fellowship and scholarship from VISTEC.

**Note and References**

## Supporting Information for:

# Engineering in-plane anisotropy in 2D materials via surface-bound ligands

Tomoaki Sakurada,[1,2,3§] Woo Seok Lee,[1,4§] Yeongsu Cho,[1] Rattapon Khamlue,[5] Petcharaphorn Chatsiri,[5] Nicholas Samulewicz,[1] Tejas Deshpande,[1] Annlin Su,[1] Peter Müller,[6] Tadashi Kawamoto,[2] Shun Omagari,[2]Martin Vacha,[2] Watcharaphol Paritmongkol,[1,5] Heather J. Kulik,[1,6] William A. Tisdale[1]*

[1]Department of Chemical Engineering, Massachusetts Institute of Technology, Cambridge, Massachusetts 02139, United States

[2]Department of Materials Science and Engineering, Institute of Science Tokyo, Ookayama 2-12-1, Meguro-ku, Tokyo 152-8552, Japan

[3]Material Integration Laboratories, AGC Inc., Yokohama, Kanagawa 230-0045, Japan

[4]Department of Materials Science and Engineering, Massachusetts Institute of Technology, Cambridge, Massachusetts 02139, United States

[5]Department of Chemistry, Massachusetts Institute of Technology, Cambridge, Massachusetts 02139, United States

[§]These authors contributed equally.

**Table of Content**

# 1. Methods

## 1.1. Chemicals

Solvents and reagents were purchased from TCI, Millipore Sigma, or Fisher Scientific and used without any further purification.

## 1.2. Equipment

### 1.2.1. Nuclear magnetic resonance (NMR) measurement

NMR spectra were recorded with a Bruker Advance 400 MHz spectrometer at 298 K. The spectral data are reported as chemical shift (in ppm). $^{1}$H-NMR was recorded with a fluorine-decoupled mode, $^{19}$F-NMR was recorded with a proton-decoupled mode, and $^{13}$C-NMR spectra were recorded with both proton- and fluorine-decoupled mode. Chemical shifts were calibrated against peak of reference chemicals as internal standard ($^{1}$H-NMR: Tetramethylsilane $\delta$= 0 ppm, $^{19}$F-NMR: 1,4-bis(trifluoromethylbenzene), $\delta = -63$ ppm) or a solvent ($^{13}$C-NMR: Chloroform $\delta$ = 77.16 ppm).

### 1.2.2. High resolution mass spectroscopy (HRMS)

HRMS was collected using a Jeol AccuTOFDART™ mass spectrometer using DART source ionization.

### 1.2.3. Thermogravimetric analysis (TGA)

TGA data was collected with a TGA Q500 differential thermal analyzer. The samples were heated from room temperature to 500 °C with a heating rate of 10°C min$^{-1}$ under $N_2$ stream (40 mL min$^{-1}$).

### 1.2.4. Powder X-ray Diffraction (PXRD)

Powder X-ray diffraction data were collected using a Rigaku SmartLab X-ray diffractometer (Cu $K_\alpha$ radiation, $\lambda$ = 1.54184 Å). A 0.04 rad Soller slit, a 2° antiscatter slit, a 10 mm mask, and a programmable divergence slit with an illuminated length of 6 mm were used in the incident beam path. The diffracted beam optics included a 0.04 rad Soller slit, a Ni filter, and an automatic receiving slit.

### 1.2.5. Fourier Transform Infrared (FT-IR) Spectroscopy

FT-IR spectra of organodiselenides and MOCs were acquired using a spectrometer and analyzed utilizing OMNIC software. The solid samples underwent scanning in the wavenumber range of 4000 to 700 cm$^{-1}$ at a resolution of 1 cm$^{-1}$, with 64 scans per spectrum conducted under a $N_2$ atmosphere.

### 1.2.6. Single-Crystal X-ray Diffraction

*For* $\mathbf{F_2(2,3)}$

X-ray diffraction measurements were performed using a Rigaku AFC7R four-circle diffractometer with graphite monochromated Mo $K_\alpha$ radiation and a rotating anode generator ($\lambda$ = 0.71069 Å). The structure was solved by the dual-space methods using SHELXT[1] and refined using the full-matrix least-squares procedure (SHELXL).[2] Anisotropic thermal parameters were adopted for all non-hydrogen atoms.

*For* $\mathbf{F_2(2,4)}$ *and* $\mathbf{F_2(2,5)}$

Low-temperature diffraction data were collected on Bruker-AXS X8 Kappa Duo diffractometers with *IµS* micro-sources using Mo $K_\alpha$ radiation ($\lambda$ = 0.71073 Å), coupled to a Photon 3 CPAD detector, performing $\phi$-and $\omega$-scans. The structures were solved by dual-space methods using SHELXT[1] and refined against $F^2$ on all data by full-matrix least squares with SHELXL-2017[2] following established refinement strategies.[3] All non-hydrogen atoms were refined anisotropically. All hydrogen atoms were included into the model at geometrically calculated positions and refined using a riding model. The isotropic displacement parameters, $U_{iso}$, of all hydrogen atoms were fixed to 1.2 times the $U_{eq}$-value of the atoms they are linked to.

The difluorophenyl rings in both structures were refined as disordered over two positions. The disorders were refined with the help of similarity restraints on 1-2 and 1-3 distances and displacement parameters as well as rigid bond restraints for anisotropic displacement parameters. In addition, the phenyl rings were restrained to be approximately planar.

Compound $\mathbf{F_2(2,5)}$ crystallizes in the triclinic centrosymmetric space group *P*-1 with one monomeric subunit in the asymmetric unit. $\mathbf{F_2(2,4)}$ crystallizes in the same space group; however, with two crystallographically independent subunits in the asymmetric unit.

*For* $\mathbf{Cl_2(2,3)}$ *and (***6***)*

X-ray diffraction data were collected on a Bruker AXS D8 Venture diffractometer equipped with an *IµS* micro-source and a Photon 3 CPAD detector, using Mo $K_\alpha$ radiation ($\lambda$ = 0.71073 Å) for $\mathbf{Cl_2(2,3)}$. and Cu $K_\alpha$ radiation ($\lambda$ = 1.5406 Å) for (**6**). The crystal was kept at a steady $T$ = 100.00 K during data collection. Data acquisition was performed using $\varphi$ and $\omega$ scan strategies. The crystal structures were processed in APEX5 and solved in the OLEX2[4] program using dual-space methods with SHELXT,[1] and refined against $F^2$ using full-matrix least-squares methods with SHELXL-2017.[2] All non-hydrogen atoms were refined anisotropically, while hydrogen atoms were placed in geometrically calculated positions and refined using a riding model.

*For (**1**), (**2**), (**4**), and (**5**)*

X-ray diffraction data were collected using a using XtaLAB Mini II (Rigaku) diffractometer using Mo-$K\alpha$ radiation ($\lambda$ = 0.71073 Å). The crystal was kept at a steady $T$ = 113.15 K during data collection. The structure was solved with the ShelXT[1] structure solution program using the Intrinsic Phasing solution method and by using Olex2[4] as the graphical interface. The model was refined with version 2018/3 of ShelXL 2018/3 using Least Squares minimisation.[2]

### 1.2.7. Optical microscopy and polarized optical microscopy.

The samples on transparent glass coverslips were mounted on an inverted microscope (Nikon, Ti-U Eclipse). Above and below the sample, a polarizer and an analyzer were placed, respectively, oriented orthogonally to each other. The sample was illuminated by an overhead light source (Nikon D-LH Halogen 12V 100W). The transmitted light through the sample was collected with an objective lens (Nikon, CFI S Plan Fluor ELWD, 40×, 0.6 NA) and then directed into a color CMOS camera (Thorlabs, DCC1645C-HQ). Polarized optical images were taken by rotating the sample stage.

### 1.2.8. Micro-absorption spectroscopy.

The samples on transparent glass coverslips were mounted on the inverted microscope. The sample was illuminated by an overhead light source (Nikon D-LH Halogen 12 V 100 W). The transmitted light through the sample was collected with the 40× objective lens and then spatially filtered through a 600 μm in diameter pinhole (spatial resolution: ~15 μm) to select a region of interest. The spatially filtered light was then directed into a spectrograph (Princeton Instruments, SP-2500) equipped with a cooled charge-coupled device (CCD) detector (Princeton Instruments, PIMAX 4: 1024 EMB). The absorbance [$A = -log_{10}(\frac{I}{I_o})$] of the sample was calculated by comparing the spectrum of the transmitted light through the crystal (I) with the spectrum of the transmitted light through the bare substrate ($I_o$) under the same experimental condition, which ensures the cancellation of polarization effects from the optics in the data. For polarization-resolved micro-absorption measurement, a linear polarizer (Thorlabs, LPVISE100-A) was inserted between the sample and the light source, and then rotated from 0° to 360° at a step size of 10°.

### 1.2.9. Micro-photoluminescence spectroscopy.

The samples on transparent glass coverslips were mounted on the inverted microscope and excited by focusing a 405 nm light (Picoquant, LDHDC- 405M, continuous wave mode) through the 40× objective lens to ~1 μm spot. The polarization 213 state of the excitation light was controlled by a circular polarizer (Thorlabs, CP1R405). After excitation, the PL was collected in the epi-configuration and passed through a dichroic mirror and a long-pass filter. It was then directed into the spectrograph with the CCD detector. For polarization-resolved PL measurement, a linear polarizer (Thorlabs, LPVISE100-A) was placed in the PL collection path and rotated from 0° to 360° at a step size of 20°. All spectra underwent Jacobian transformation from wavelength to photon energy,[5] but have not been corrected for wavelength-dependent efficiency of the spectrograph or CCD camera. We confirmed that the polarization response from the optical components in the PL collection path is negligible by conducting measurements using unpolarized broadband light (Thorlabs, MCWHL2-C3) placed on the sample stage.

### 1.2.10. Time-Resolved PL Spectroscopy

Time-resolved PL measurements were performed using the same microscope setup as steady-state PL spectroscopy with some modifications. A variable repetition-rate 405 nm pulsed laser diode (Picoquant, LDHDC-405M) was used as the excitation light source. The detector used was a Si avalanche photodiode (Micro Photon Devices) connected to a counting board for time-correlated single-photon counting (PicoQuant, PicoHarp 300).

### 1.2.11. Temperature-Dependent PL Spectroscopy.

Temperature dependent photoluminescence (PL) spectroscopy was performed by mounting samples in a microscopy cryostat (Janis Research, ST-500) and flowing liquid nitrogen through a coldfinger attached to the base of the cryostat.

### 1.2.12. Photoluminescent Quantum Yield.

The measurement of PL quantum yield (QY) was performed at room temperature using the absolute quantum yield method in an integrating sphere.[6] A colloidal dispersion of MOCs was prepared by sonication of crystals in 2-propanol (*ca.* 0.5 wt%) and then placed into a quartz cuvette with a 10 mm path length. The excitation light from a 405 nm laser diode (Picoquant, LDHDC-405M, continuous wave mode) was directed into an integrating sphere (Labsphere) containing the sample. The output signal was collected by an optical fiber mounted at an exit port of the integrating sphere and directed to a fiber-optic spectrometer (Avantes).

### 1.2.13. Density functional theory Calculations.

Density functional theory (DFT) calculations were performed using Vienna Ab initio Simulation Package (VASP) version 6.3.1.[7–10] The PBE functional[11] and projector augmented wave pseudopotentials[12] were employed with a kinetic energy cutoff of 1000 eV. Dispersion interactions were treated using the DFT-D3 method[13] with Becke-Johnson damping.[14] Single crystal X-ray diffraction analysis provided two possible positions for the phenyl rings in **$F_2$(2,4)** and **$F_2$(2,5)**, thus atomic coordinates with higher occupancy were selected for the calculations. Atomic positions were optimized while keeping the cell parameters fixed, using the measured structure as the starting point. Structural optimization used a 3×3×1 k-mesh and proceeded until the total energy difference between steps dropped below 1 meV. Self-consistent field calculations were performed on the optimized geometry with a 6×6×1 k-mesh to obtain the charge density and the projected density of states. The band structure was then calculated from the charge density generated with this denser k-point mesh. The effective mass was estimated with the Kane quasi-linear dispersion to capture the non-parabolic bands, using the Effmass package.[15] Interaction energies between two ligands (i.e., benzene for **AgSePh**, 1,2-difluorobenzene for **$F_2$(2,3)**, and 1,2-dichlorobenzene for **$Cl_2$(2,3)**) were calculated by placing the pair of two ligands in a 40 Å cubic cell to avoid interactions with periodic images. All atoms were positioned within the xz plane, and the in-plane orientation relative to the x axis matched the observed orientation in the corresponding MOC in order to reproduce the ligand-ligand interactions present in the MOC. Then these ligands were systematically separated in 0.1 Å increments, and single point energies were evaluated.

## 2. Synthesis

### 2.1. Synthesis of Diselenides

All diselenides were synthesized by a reaction of Grignard reagent and elemental selenium. Here, a typical synthesis procedure is shown for 1,2-bis(2,3-difluorophenyl) diselenide as an example. Due to high crystallinity of diselenides the structure was determined by single-crystal XRD except for (**3**).

*Synthesis of 1,2-bis(2,3-difluorophenyl) diselenide (**1**)*

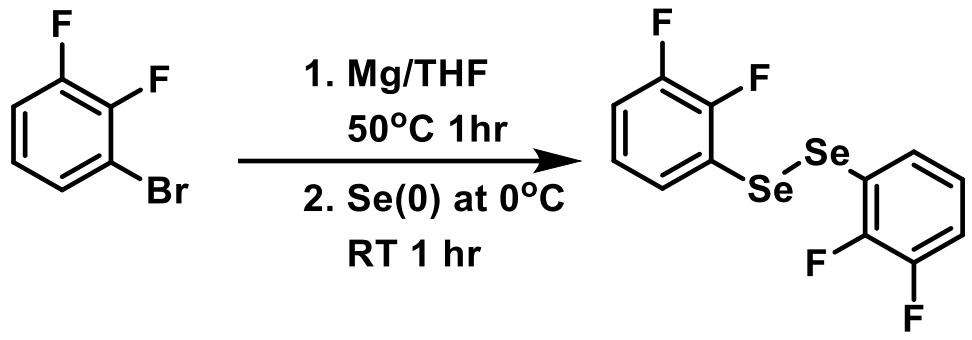

To a dispersion of Mg (2.52 g, 110 mmol) and 60 mL of anhydrous tetrahydrofuran (THF), a small amount of 1-bromo-2,3-difluorobenzene (1.9 g, 10 mmol) was added and stirred under $N_2$ atmosphere for 20 min. Once the internal temperature increased, a second portion of 1-bromo-2,3-difluorobenzene (17.4 g, 90 mmol) in THF (40 mL) was added dropwise to a vigorously stirred dispersion over 15 min. The reaction mixture was heated to 50 °C and stirred for 1 h, before being cooled in an ice bath. Elemental selenium (8.8 g, 110 mmol) was added in a single portion, and the reaction mixture was allowed to warm to room temperature and stirred for 1 h. The mixture was filtered thorough Celite left at ambient condition overnight. After that, the solvent was evaporated, and the obtained residue was redissolved in dichloromethane. This organic phase was washed sequentially with aqueous ammonium chloride solution and brine, then dried over sodium sulfate. Solvent was evaporated again under reduced pressure using a rotary evaporator, and the crude product was purified by column chromatography (using hexanes as the eluent) to afford 1,2-bis(2,3-difluorophenyl) diselenide as orange oil (*ca.* 10g). Orange oil transformed into yellow crystals when stored at ambirnt conditions for 1 week.

$^{1}$H-NMR (400 MHz, $CDCl_3$) $\delta$ 7.39 (dd, $J$ = 7.9, 1.7 Hz, 2H), 7.12 (dd, $J$ = 8.2, 1.5 Hz, 2H), 7.03 (t, $J$ = 8.1 Hz, 2H). $^{13}$C-NMR (101 MHz, $CDCl_3$) $\delta$ 150.3, 149.3, 129.1, 124.9, 119.1, 117.8. $^{19}$F-NMR (376 MHz, $CDCl_3$) $\delta$ −126.9 (d, $J$ = 23.6 Hz, 4F), −136.2 (d, $J$ = 23.0 Hz, 4F).

HRMS calculated for $C_{12}H_6F_4Se_2$ $[M]^+$ 385.8733, found 385.8762.

*Synthesis of 1,2-bis(2,4-difluorophenyl) diselenide (**2**)*

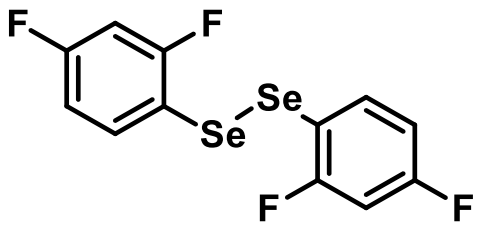

yellow crystals.

$^{1}$H-NMR (400 MHz, $CDCl_3$) $\delta$ 7.59 (d, $J$ = 8.1 Hz, 2H), 6.86-6.83 (m, 4H). $^{13}$C-NMR (101 MHz, $CDCl_3$) $\delta$ 164.0, 161.8, 136.6, 112.5, 112.3, 104.4. $^{19}$F-NMR (376 MHz, $CDCl_3$) $\delta$ −96.9 (d, $J$ = 9.0 Hz, 2F), −108.0 (d, $J$ = 9.0 Hz, 2F)

*Synthesis of 1,2-bis(2,5-difluorophenyl) diselenide (**3**)*

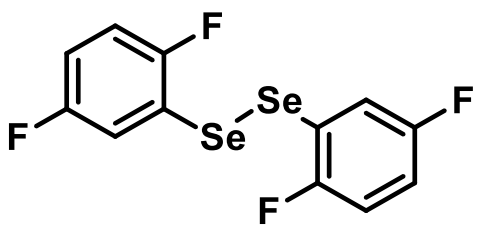

yellow crystals.

$^{1}$H-NMR (400 MHz, $CDCl_3$) $\delta$ 7.34 (d, $J$ = 2.9 Hz, 2H), 7.01 (d, $J$ = 9.1 Hz, 2H), 6.94 (dd, $J$ = 8.8, 2.9 Hz, 2H). $^{13}$C-NMR (101 MHz, $CDCl_3$) $\delta$ 159.2, 156.9, 119.8, 117.9, 116.6, 116.4. $^{19}$F-NMR (376 MHz, $CDCl_3$) $\delta$ −110.2 (d, $J$ = 16.8 Hz, 2F), −117.0 (d, $J$ = 16.8 Hz, 2F)

*Synthesis of 1,2-bis(2,3,4-trifluorophenyl) diselenide (**4**)*

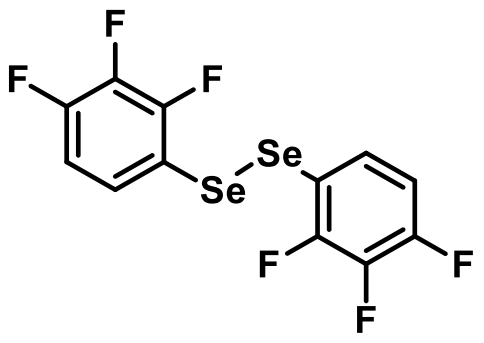

yellow crystals.

$^{1}$H-NMR (400 MHz, $CDCl_3$) $\delta$ 7.34 (d, $J$ = 9.3 Hz, 2H), 6.95 (d, $J$ = 8.8 Hz, 2H). $^{13}$C-NMR (101 MHz, $CDCl_3$) $\delta$ 152.3, 150.9, 140.0, 129.2, 113.6, 113.2. $^{19}$F-NMR (376 MHz, $CDCl_3$) $\delta$ −120.4 (dd, $J$ = 22.7, 8.7 Hz, 2F), −131.3 (q, $J$ = 9.5 Hz, 2F), −157.4 (dd, $J$ = 23.0, 20.2 Hz, 2F)

*Synthesis of 1,2-bis(2,4,5-trifluorophenyl) diselenide (**5**)*

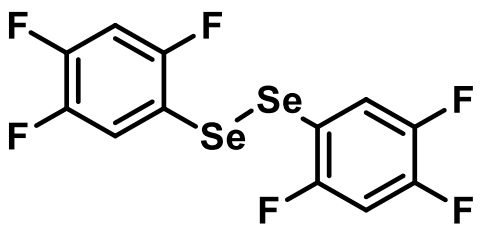

yellow crystals.

$^{1}$H-NMR (400 MHz, $CDCl_3$) $\delta$ 7.45 (s, 2H), 6.94 (s, 2H). $^{13}$C-NMR (101 MHz, $CDCl_3$) $\delta$ 156.6, 151.0, 147.3, 122.4, 111.9, 105.9. $^{19}$F-NMR (376 MHz, $CDCl_3$) $\delta$ −103.5 (dd, $J$ = 14.0, 5.0 Hz, 2F), −131.2 (dd, $J$ = 21.0, 4.8 Hz, 2F), −140.4 (dd, $J$ = 20.8, 14.0 Hz, 2F)

*Synthesis of 1,2-bis(2,3-dichlorophenyl) diselenide (**6**)*

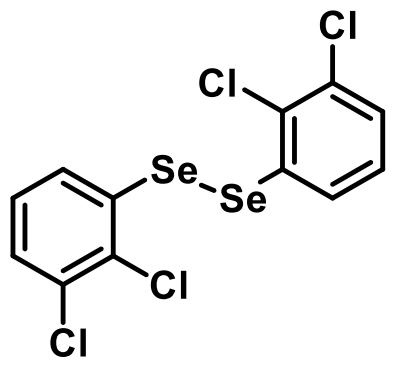

Pale yellow crystals.

$^{1}$H-NMR (400 MHz, $CDCl_3$) $\delta$ 7.47 (dd, $J$ = 8.0, 1.3 Hz, 2H), 7.35 (dd, $J$ = 8.0, 1.3 Hz, 2H), 7.10 (t, $J$ = 8.0 Hz, 2H). $^{13}$C-NMR (101 MHz, $CDCl_3$) δ 133.1, 131.1, 130.7, 128.9, 128.4, 128.2.

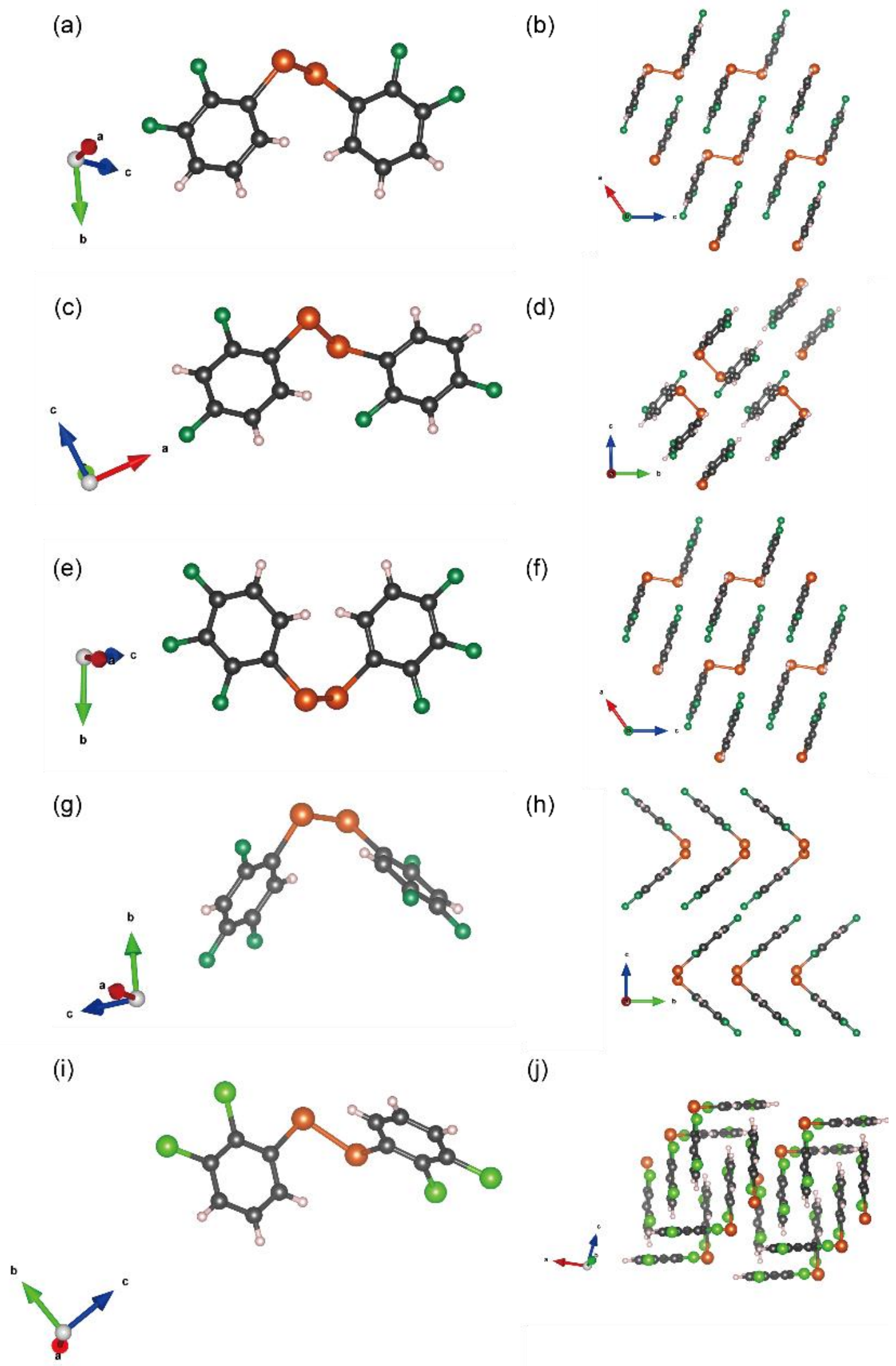

**Figure S1.** Molecular structure and molecular assembly in crystals of (**1**) (a,b), (**2**) (c,d), (**4**) (e,f), (**5**) (g,h), and (**6**) (i,j). C, Se, F, Cl and H atoms are depicted by black, orange, green, light green and pink.

**Table S1.** Crystal Data and Structure Refinement for Organodiselenides

| | (**1**) | (**2**) | (**4**) | (**5**) | (**6**) |
|---|---|---|---|---|---|
| CCDC Deposition No. | 2501947 | 2501948 | 2501952 | 2501953 | 2473793 |
| Empirical formula | $C_{12}H_6F_4Se_2$ | $C_{12}H_6F_4Se_2$ | $C_6H_2F_3Se$ | $C_6H_2F_3Se$ | $C_{24}H_{12}Cl_8Se_4$ |
| $M_r$ | 384.09 | 384.09 | 210.04 | 210.04 | 899.78 |
| Temperature (K) | 113(2) | 113(2) | 113(2) | 113(2) | 100(2) |
| Wavelength (Å) | 0.71073 | 0.71073 | 0.71073 | 0.71073 | 1.54184 |
| Crystal system | Monoclinic | Triclinic | Monoclinic | Monoclinic | Triclinic |
| Space group | *C*2/*c* | *P*-1 | *C*2/*c* | *C*2/*c* | *P*-1 |
| *a* (Å) | 15.792(3) | 7.6301(4) | 16.246(2) | 6.9498(7) | 7.6786(2) |
| *b* (Å) | 7.0821(9) | 7.8493(6) | 7.3056(6) | 8.9884(7) | 7.6891(2) |
| *c* (Å) | 12.588(2) | 9.8423(6) | 12.4165(18) | 19.4288(16) | 12.9010(4) |
| α (°) | 90 | 91.197(6) | 90 | 90 | 92.8450(10) |
| β (°) | 124.75(3) | 90.665(5) | 125.33(2) | 100.160(8) | 91.5480(10) |
| γ (°) | 90 | 98.454(6) | 90 | 90 | 114.1300(10) |
| $V$ (Å$^3$) | 1156.8(5) | 582.87(7) | 1202.2(4) | 1194.64(18) | 693.35(3) |
| $Z$ | 4 | 2 | 8 | 8 | 1 |
| Calculated density (Mg/m$^3$) | 2.205 | 2.188 | 2.321 | 2.336 | 2.155 |
| Absorption coefficient (mm$^{-1}$) | 6.420 | 6.371 | 6.212 | 6.251 | 13.570 |
| $F$(000) | 728 | 364 | 792 | 792 | 428 |
| Crystal size (mm$^3$) | 0.30 × 0.20 × 0.10 | 0.30 × 0.15 × 0.08 | 0.50 × 0.40 × 0.20 | 0.30 × 0.30 × 0.20 | 0.238 × 0.085 × 0.036 |
| $\theta$ range for data collection (°) | 3.140 to 30.652 | 2.624 to 30.690 | 3.074 to 30.637 | 3.743 to 30.597 | 3.434 to 66.621 |
| Index ranges | $-22 \le h \le 14$, $-9 \le k \le 10$, $-16 \le l \le 17$ | $-10 \le h \le 10$, $-11 \le k \le 11$, $-13 \le l \le 13$ | $-21 \le h \le 23$, $-10 \le k \le 10$, $-17 \le l \le 17$ | $-9 \le h \le 9$, $-12 \le k \le 8$, $-26 \le l \le 27$ | $-9 \le h \le 9$, $-9 \le k \le 8$, $-15 \le l \le 15$ |
| Reflections collected | 3929 | 9170 | 5137 | 3346 | 16780 |
| Independent reflections | 1540 [$R_{int}$ = 0.0396] | 3368 [$R_{int}$ = 0.0401] | 1784 [$R_{int}$ = 0.0572] | 1654 [$R_{int}$ = 0.0230] | 2422 [$R_{int}$ = 0.0301] |
| Completeness to $\theta$ = 25.242° | 99.9% | 99.1% | 99.8% | 97.6% | 98.9% |
| Data / restraints / parameters | 1540 / 0 / 82 | 3368 / 0 / 163 | 1784 / 0 / 91 | 1654 / 0 / 91 | 2422 / 0 / 163 |
| Goodness-of-fit on $F^2$ | 1.092 | 1.116 | 1.094 | 1.036 | 1.147 |
| Final R indices [$I > 2\sigma(I)$] | $R$1 = 0.0533, $wR$2 = 0.1172 | $R$1 = 0.0501, $wR$2 = 0.1078 | $R$1 = 0.0520, $wR$2 = 0.1137 | $R$1 = 0.0315, $wR$2 = 0.0605 | $R$1 = 0.0181, $wR$2 = 0.0445 |
| R indices (all data) | $R$1 = 0.0746, $wR$2 = 0.1236 | $R$1 = 0.0734, $wR$2 = 0.1135 | $R$1 = 0.0751, $wR$2 = 0.1212 | $R$1 = 0.0464, $wR$2 = 0.0645 | $R$1 = 0.0189, $wR$2 = 0.0446 |
| Largest diff. peak and hole (e.Å$^{-3}$) | 1.980 and −2.126 | 1.801 and −1.031 | 1.694 and −1.095 | 0.669 and −0.530 | 0.363 and −0.476 |

## 2.2. Synthesis of AgSePh and Halogenated Mithrenes

A solution of organodiselenides in 1-butylamine (10 mM, 5 mL) was mixed with a solution of $AgNO_3$ in 1-butylamine (10 mmol, 5 mL) filtered through a 0.2 µm PTFE disc syringe filter. The filtered solution (10 mL) was transferred to a 20 mL glass vial, which was then placed inside a larger 110 mL glass vial containing 20 mL deionized water. This setup allowed water to diffuse into the small vial while organic solvents gradually evaporated. The deionized water in the larger glass jar was renewed every 3 days. Crystals began to appear after approximately 6 days at room temperature. Obtained crystals were washed with 2-propanol, toluene, and 2-propanol and stored in 2-propanol.

## 3. Supplementally Figures and Tables

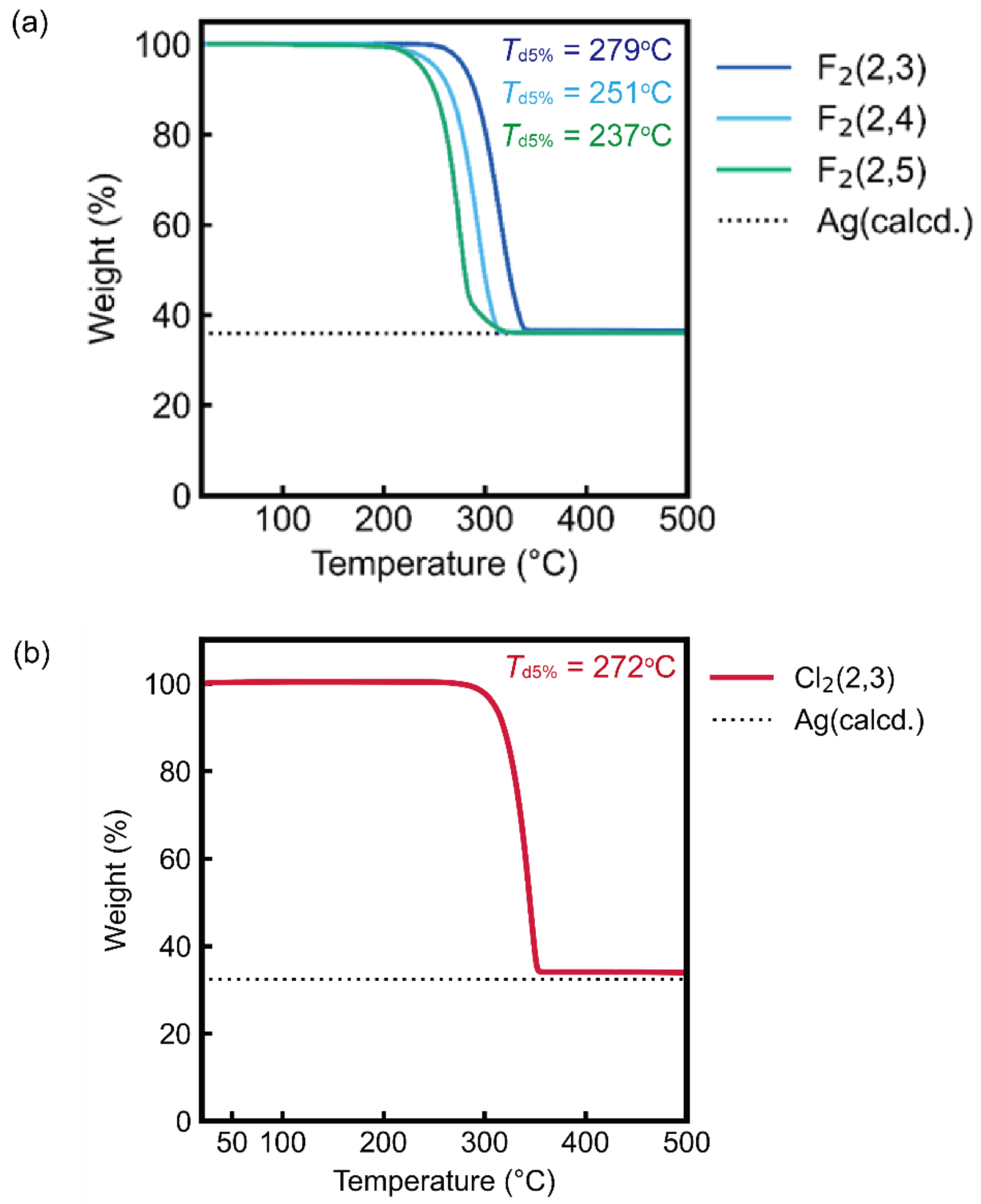

**Figure S2.** TGA curve for (a) **$F_2$(2,3)**, **$F_2$(2,4)**, and **$F_2$(2,5)** and (b) **$Cl_2$(2,3)** under $N_2$ flow.

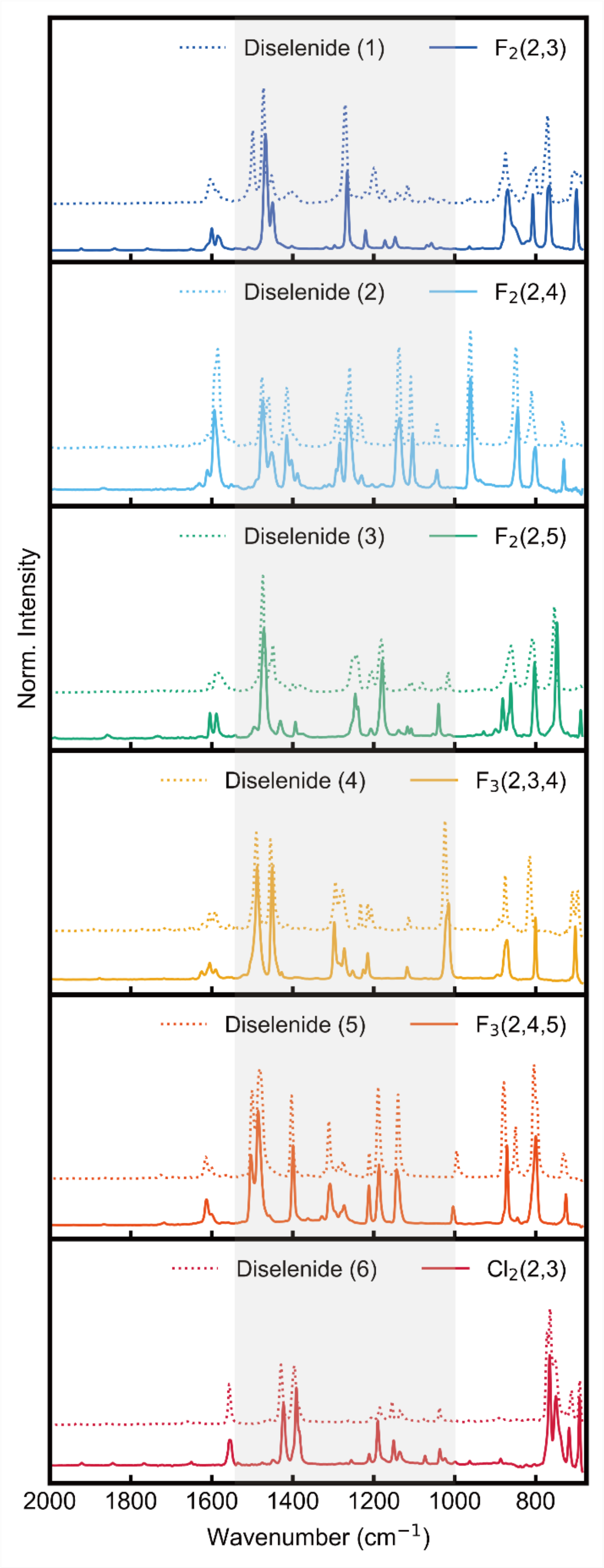

**Figure S3.** IR spectra of precursors (dotted line) and MOCs (solid line).

**Table S2.** Optical properties of MOCs at 300 K.

| MOCs | $\lambda_{em}$ nm / eV | FWHM (nm), (meV) | $\tau_{em}$ (ns) | PLQY (%) |
|---|---|---|---|---|
| **$F_2$(2,3)** | 476 / 2.60 | 16.8 / 112 | 0.5 | 0.9 |
| **$F_2$(2,4)** | 476 / 2.60 | 21.2 / 150 | 0.1 | 0.1 |
| **$F_2$(2,5)** | 472 / 2.62 | 20.2 / 108 | 0.7 | 1.2 |
| **$F_3$(2,3,4)** | 468 / 2.65 | 20.7 / 117 | 0.1 | 0.1 |
| **$F_3$(2,4,5)** | 483 / 2.57 | 24.3 / 132 | 0.1 | 0.1 |
| **AgSePh** | 467 / 2.66 | 16.1 / 91 | 0.1 | 0.1 |

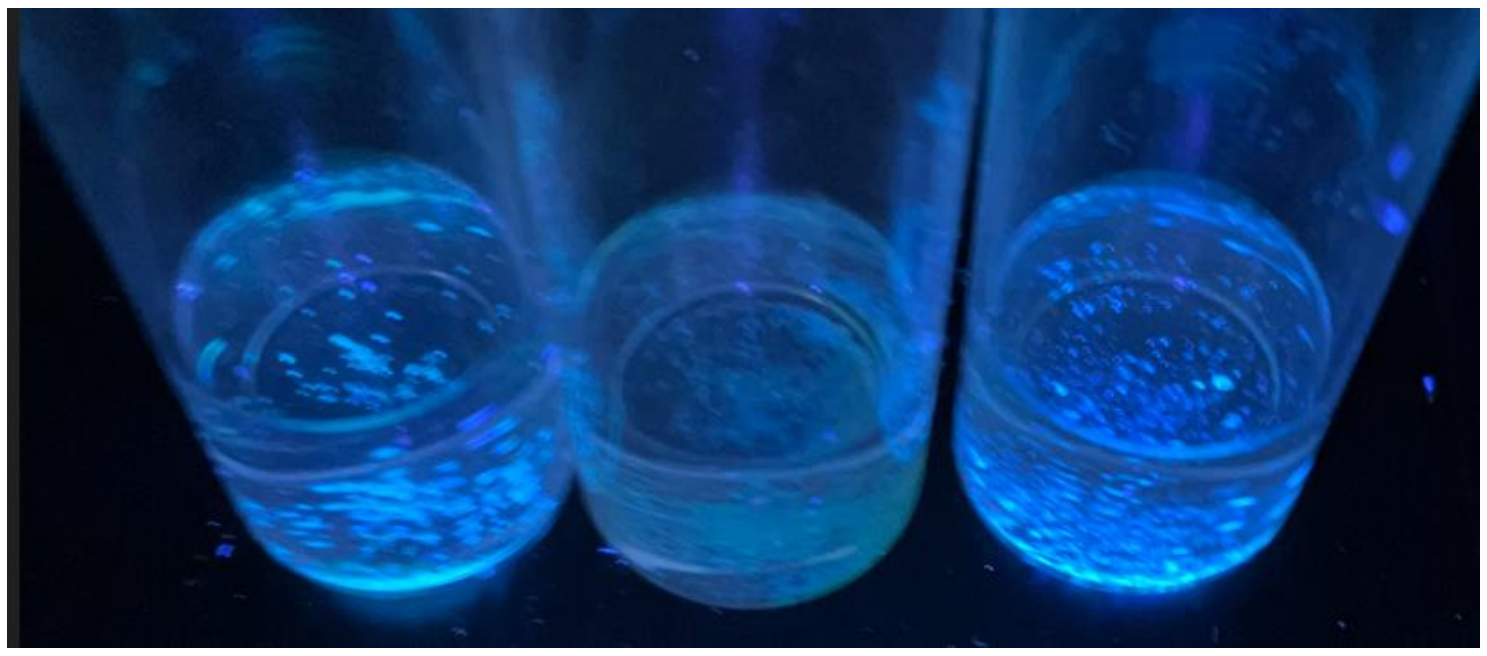

**Figure S4.** Optical image of MOC crystal dispersion (**$F_2$(2,3)**, **$F_2$(2,4)** and **$F_2$(2,5)**) in 2-propanol under UV light (365 nm) excitation.

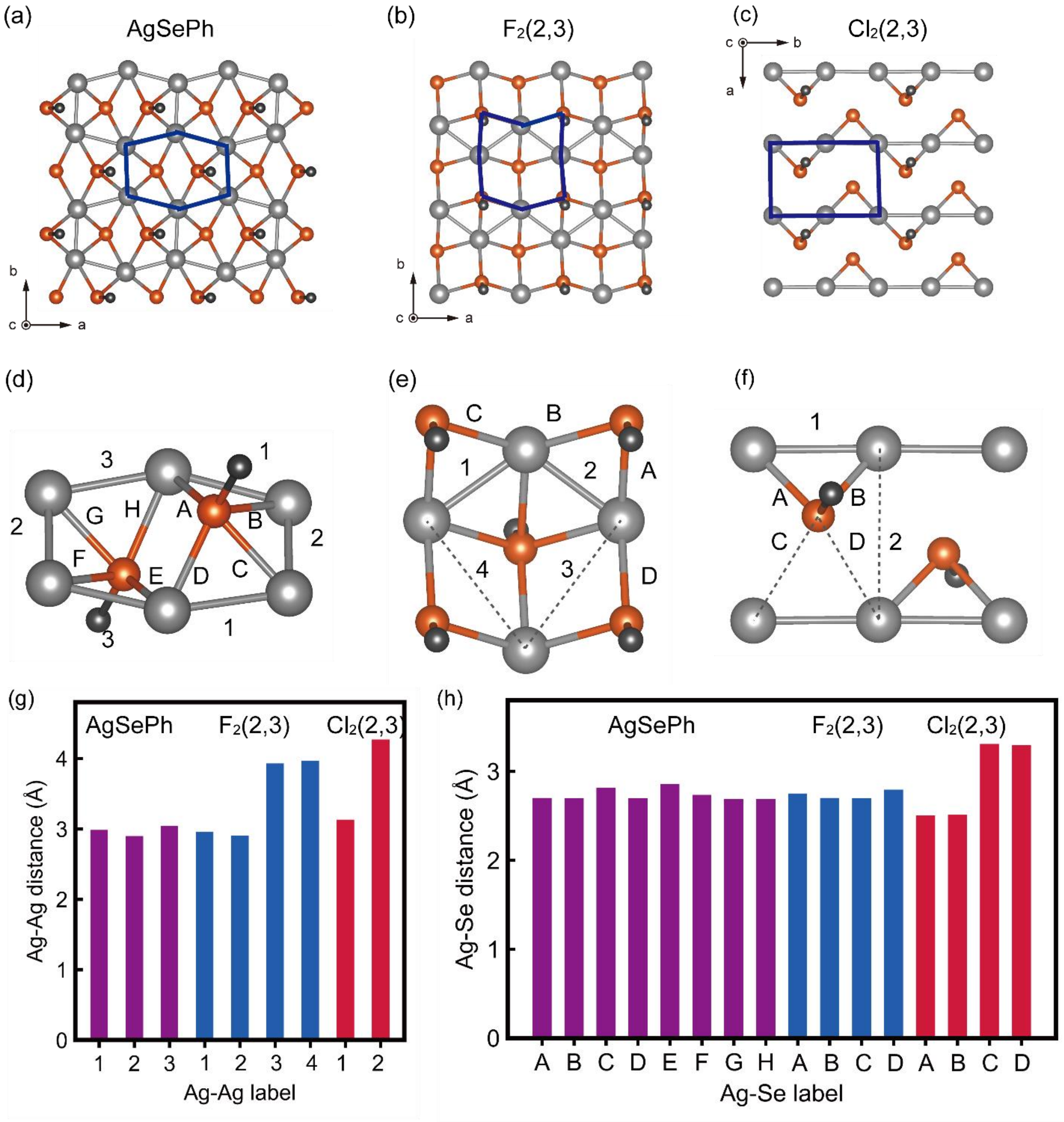

**Figure S5.** Ag-Se layer structure for (a) **AgSePh**, (b) **$F_2$(2,3)**, and (c) **$Cl_2$(2,3)**. Repeating unit of Ag-Se layer structure for (d) **AgSePh**, (e) **$F_2$(2,3)**, and (f) **$Cl_2$(2,3)**. Labels corresponding to distance of Ag-Ag and Ag-Se interactions. (g) Ag-Ag separation for each MOC. (h) Ag-Se separation for each MOC.

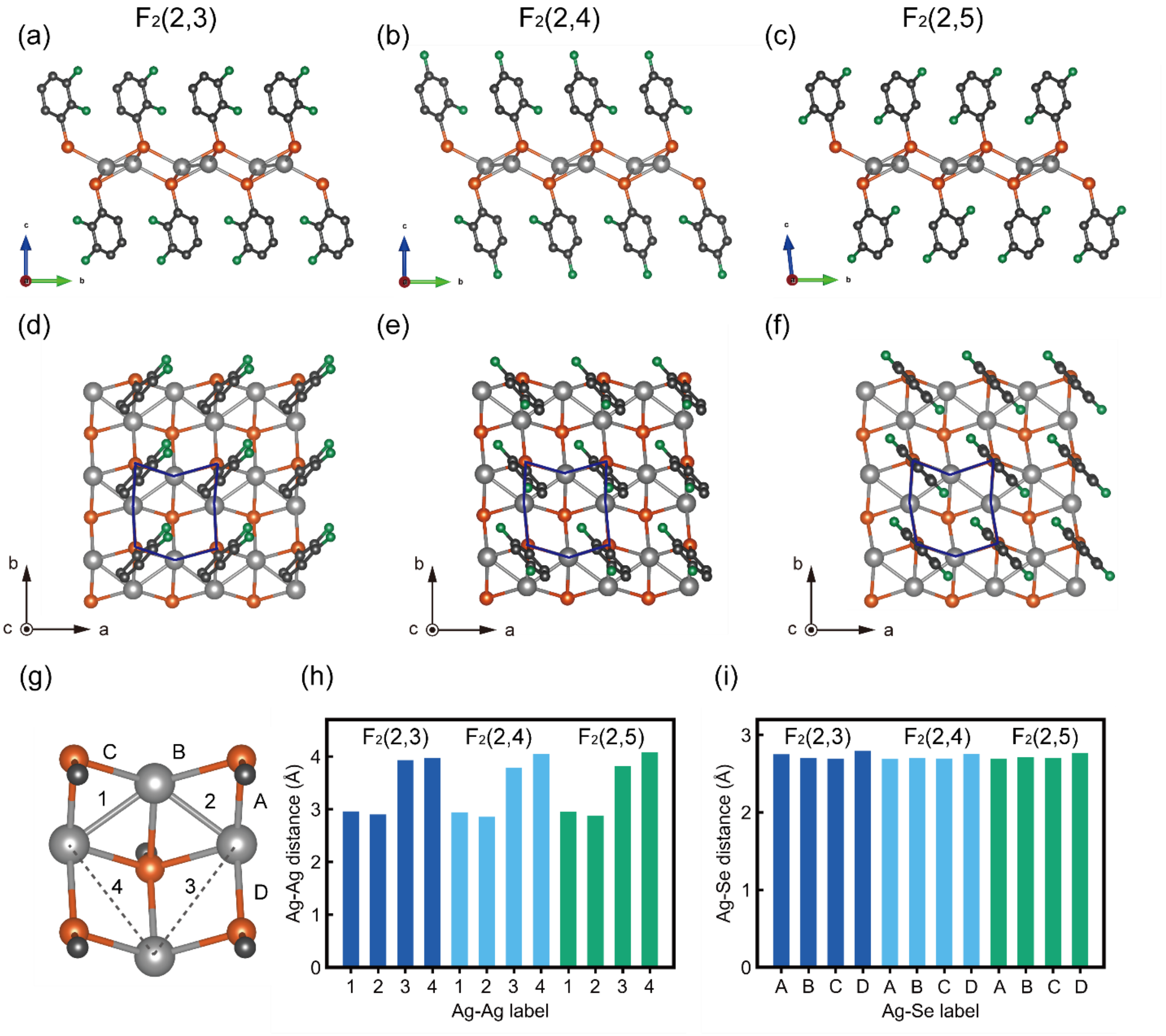

**Figure S6.** Structure for (a) $\mathbf{F_2(2,3)}$, (b) $\mathbf{F_2(2,4)}$, and (c) $\mathbf{F_2(2,5)}$ from side view. Ag-Se layer structure of (d) $\mathbf{F_2(2,3)}$, (e) $\mathbf{F_2(2,4)}$, and (f) $\mathbf{F_2(2,5)}$. (g) Repeating core of Ag-Se layer. (h) Ag-Ag separation and (i) Ag-Se separation for each MOC.

$Cl_2(2,3)$

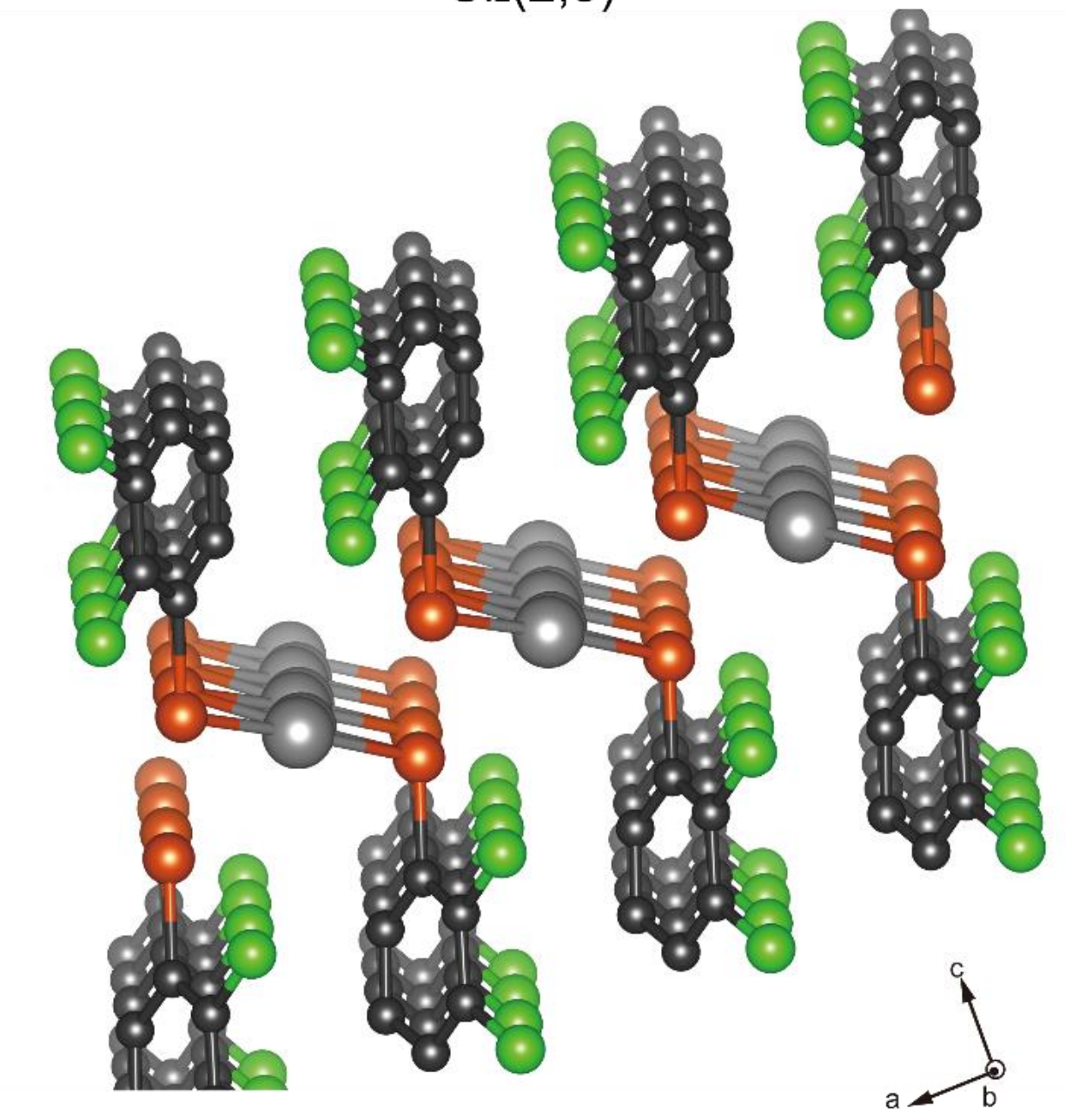

**Figure S7.** Structure of **$Cl_2$(2,3)**.

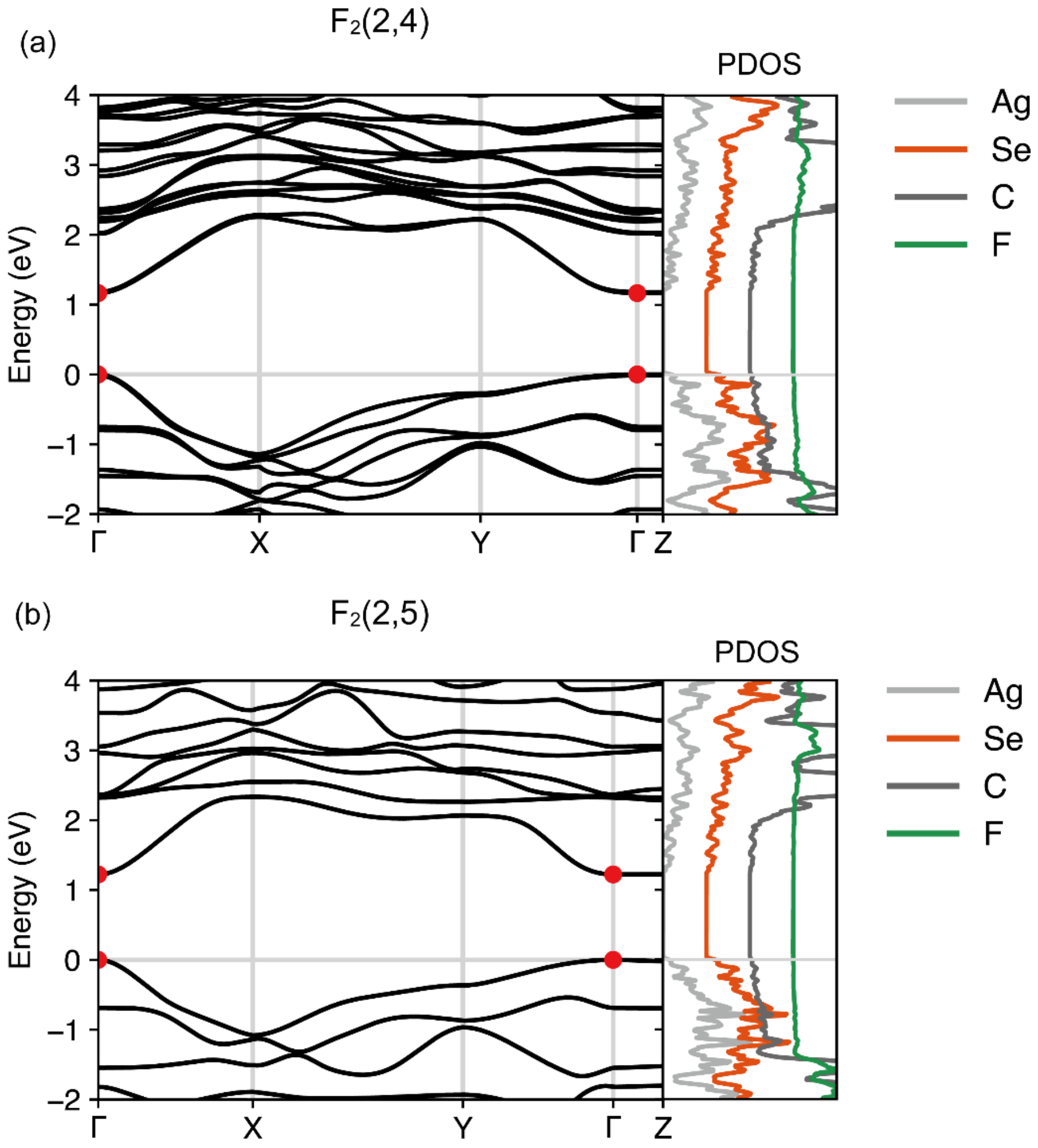

**Figure S8.** Electronic band structure and projected density of states (PDOS) of $\mathbf{F_2(2,4)}$ and $\mathbf{F_2(2,5)}$ calculated using the PBE functional. Red dots mark the locations of the valence band maximum and the conduction band minimum.

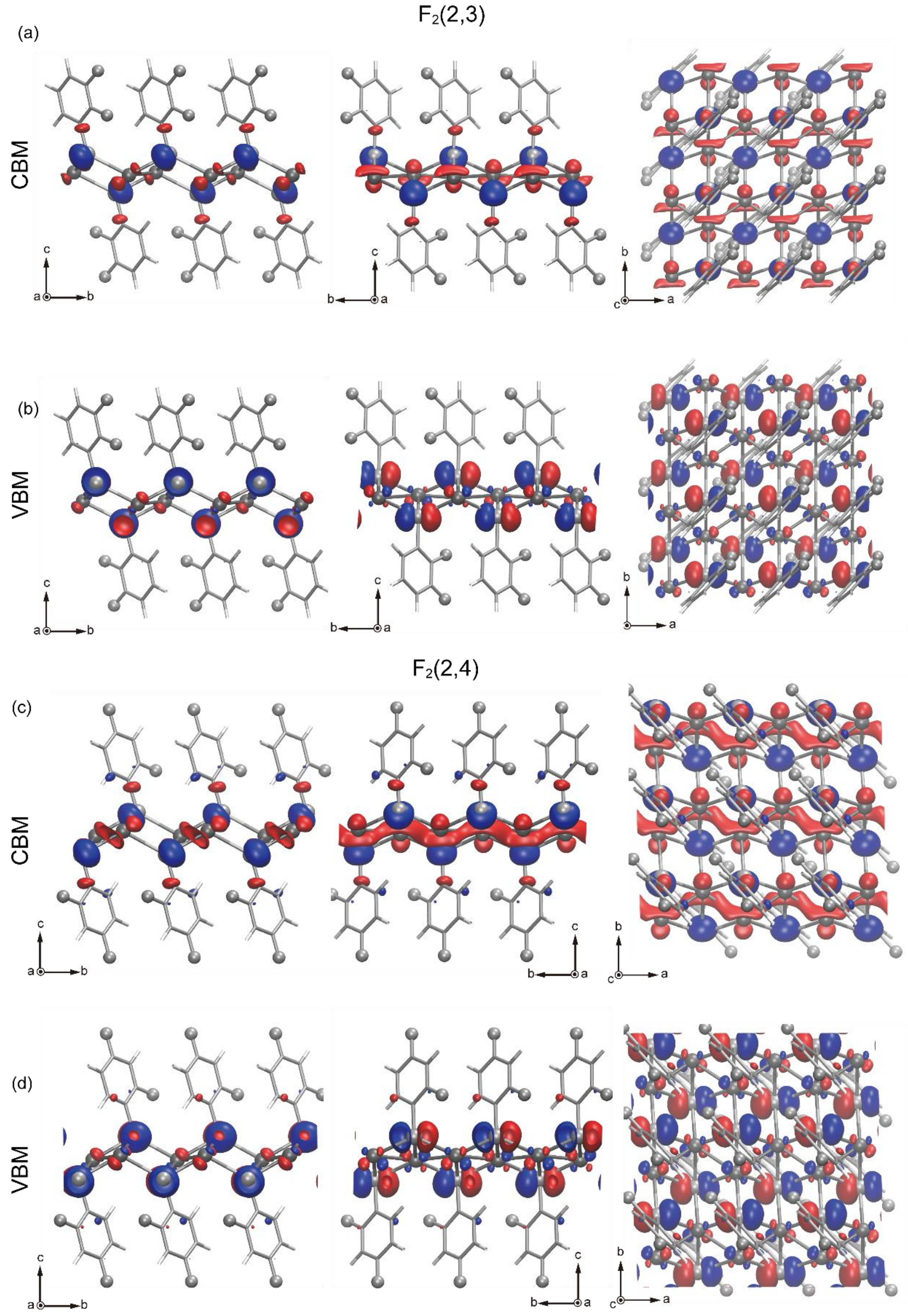

**Figure S9.** Wavefunctions for $\mathbf{F_2(2,3)}$ at (a) CBM and (b) VBM and for $\mathbf{F_2(2,4)}$ at (c) CBM and (d) VBM. C and H atoms are represented with a stick model, and grey spheres denote Ag, Se, and F. Red and blue surfaces represent the positive and negative signs of the wavefunctions at an isosurface level of 0.003 Å$^{-3}$.

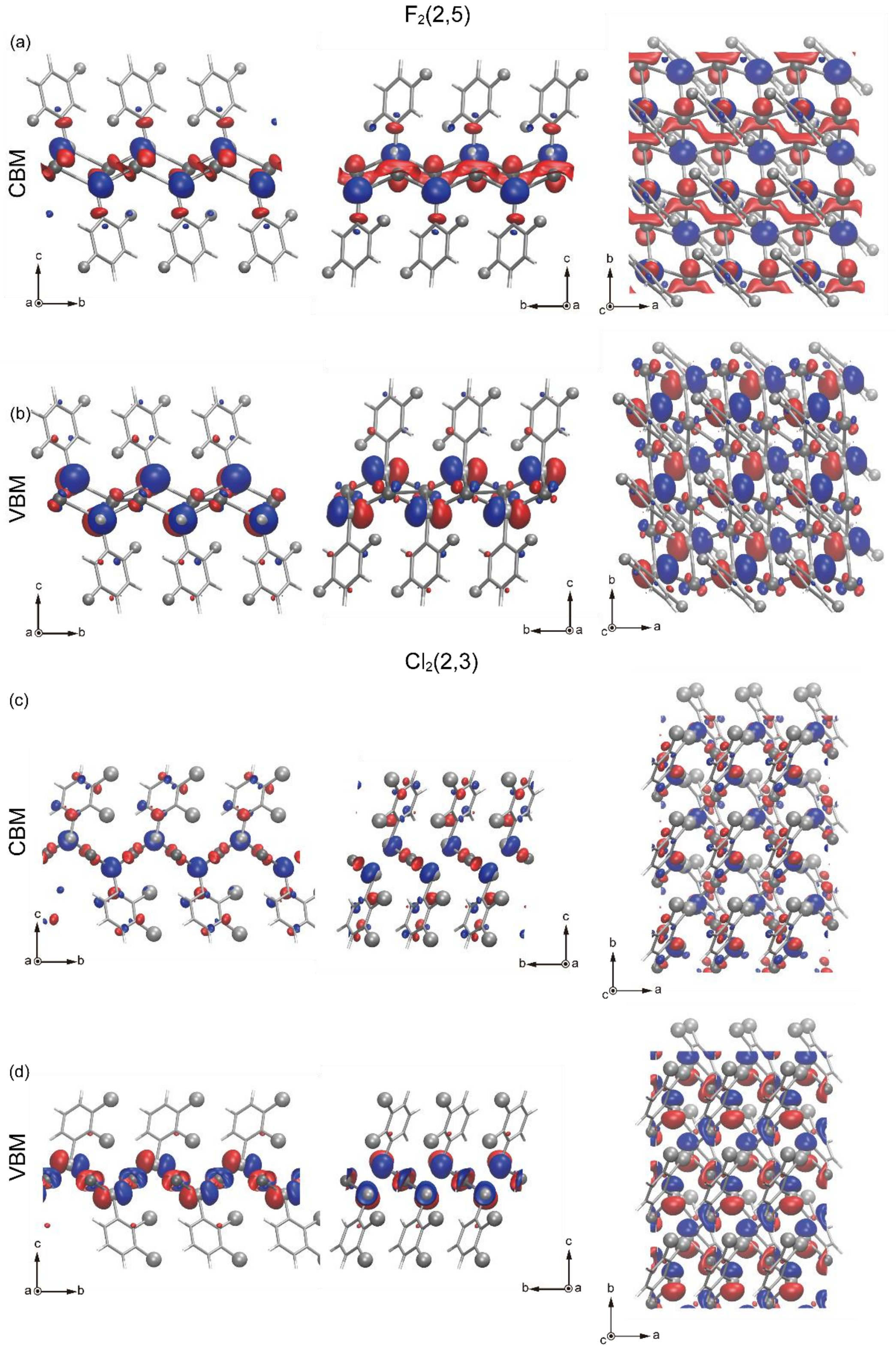

**Figure S10.** Wavefunctions for **$F_2$(2,5)** at (a) CBM and (b) VBM and for **$Cl_2$(2,3)** at (c) CBM and (d) VBM. C and H atoms are represented with a stick model, and grey spheres denote Ag, Se, F, and Cl. Red and blue surfaces represent the positive and negative signs of the wavefunctions at an isosurface level of 0.003 Å$^{-3}$.

**Table S3**. Effective masses of holes and electrons along different orientation for each MOC.

| | | **AgSePh** | **$F_2$(2,3)** | **$F_2$(2,4)** | **$F_2$(2,5)** | **$Cl_2$(2,3)** |
|---|---|---|---|---|---|---|
| Hole | $m^*_x/m_0$ | 1.14 | 0.28 | 0.27 | 0.32 | 0.90 (X→Γ) |
| | $m^*_y/m_0$ | 0.77 | 3.77 | 3.51 | 2.40 | - |
| | Ratio | 1.48 | 13.5 | 13.0 | 7.5 | - |
| Electron | $m^*_x/m_0$ | 1.28 | 0.35 | 0.37 | 0.36 | 2.17 (CBM→X) |
| | $m^*_y/m_0$ | 0.24 | 1.38 | 1.50 | 0.60 | 2.98 (CBM→Γ) |
| | Ratio | 3.05 | 3.94 | 4.05 | 1.67 | 1.37 |

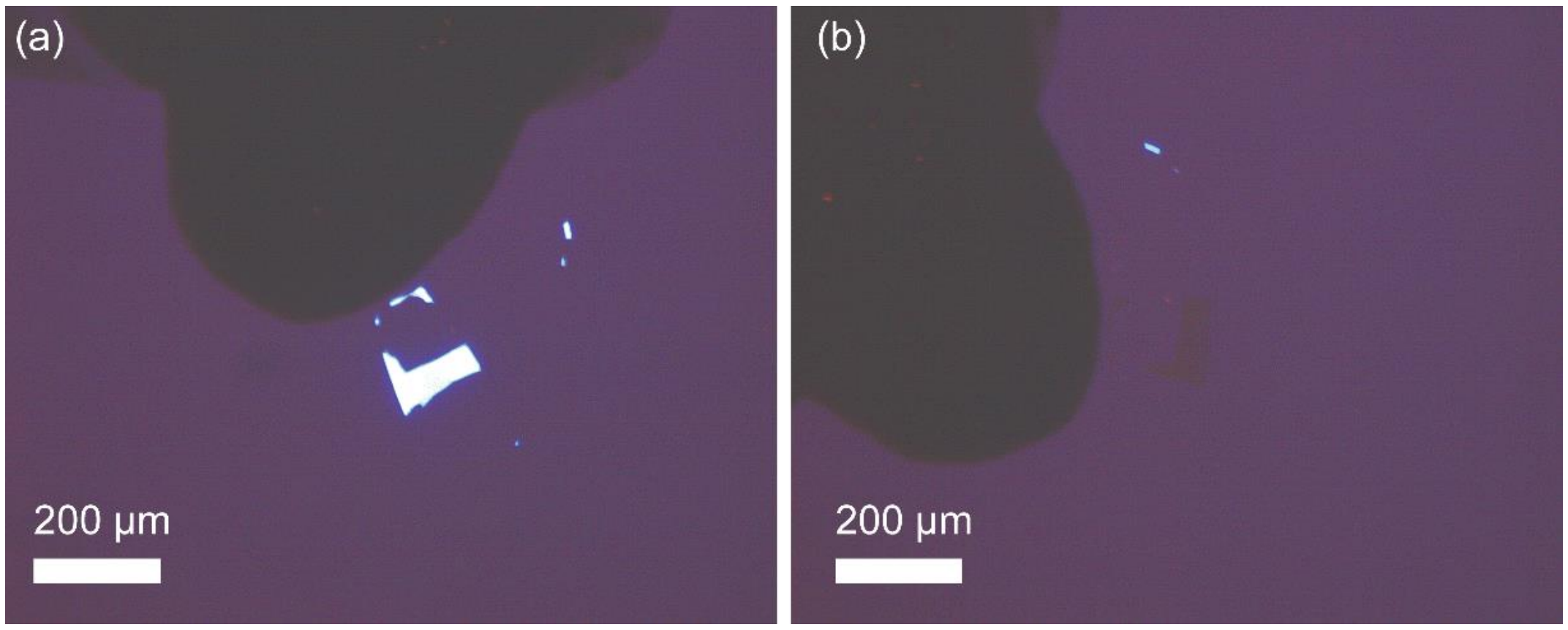

**Figure S11**. Polarized pptical micrograph of the exfoliated **$F_2$(2,3)** crystals. The change in brightness across the crystal from (a) complete brightness to (d) darkness.

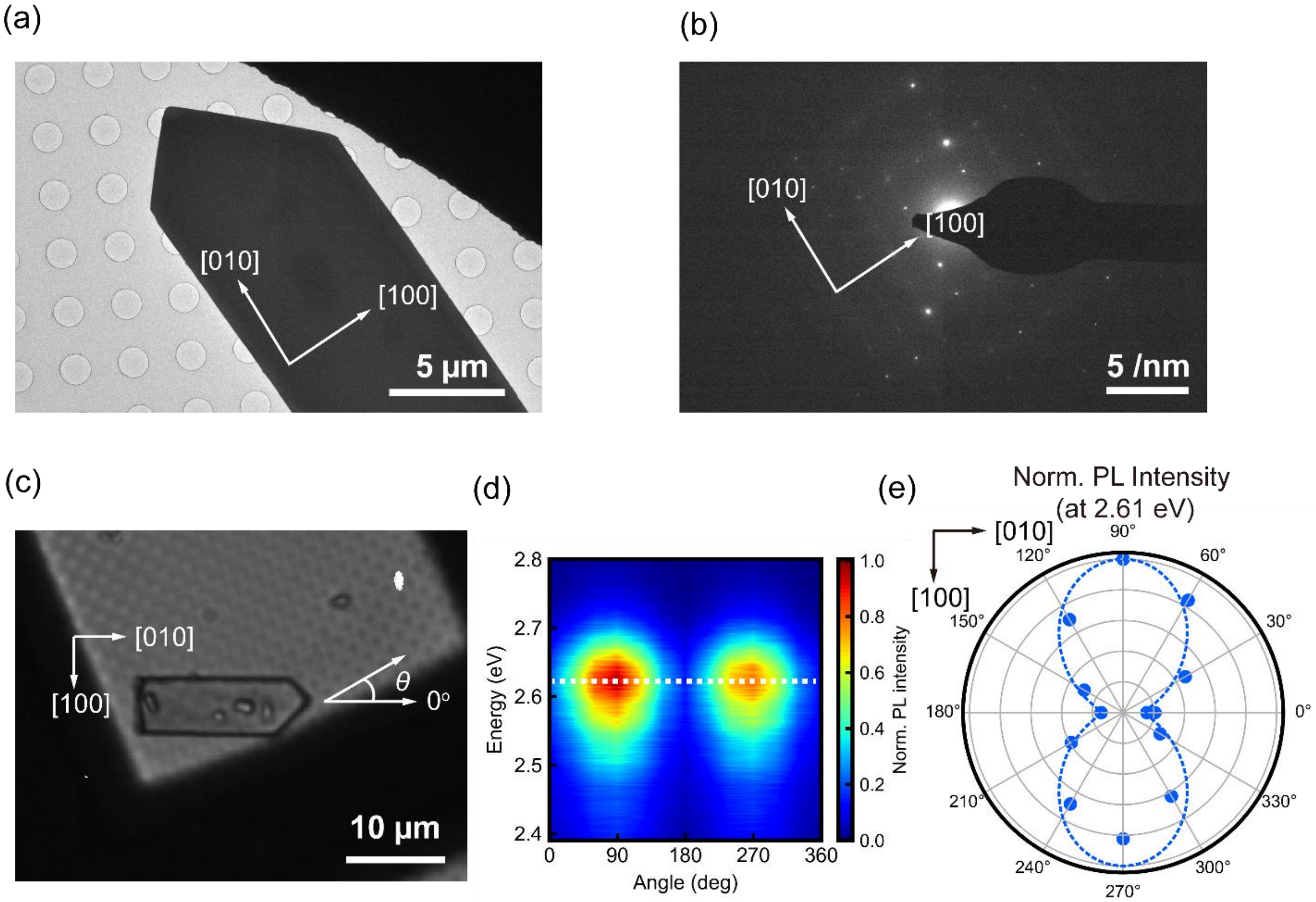

**Figure S12. Crystal Orientation and Direction of Optical Anisotropy.** (a) TEM image, (b) electron diffraction pattern, and (c) optical micrograph of the same **$F_2$(2,3)** crystal. (d) PL spectrum of the same **$F_2$(2,3)** crystal as a function of polarization angle of emitted light. The zero angle corresponds to the [010] direction as marked in (c). (e) Polar plot of normalized PL intensity at 2.61 eV, showing that the PL is polaried along the [100] direction.

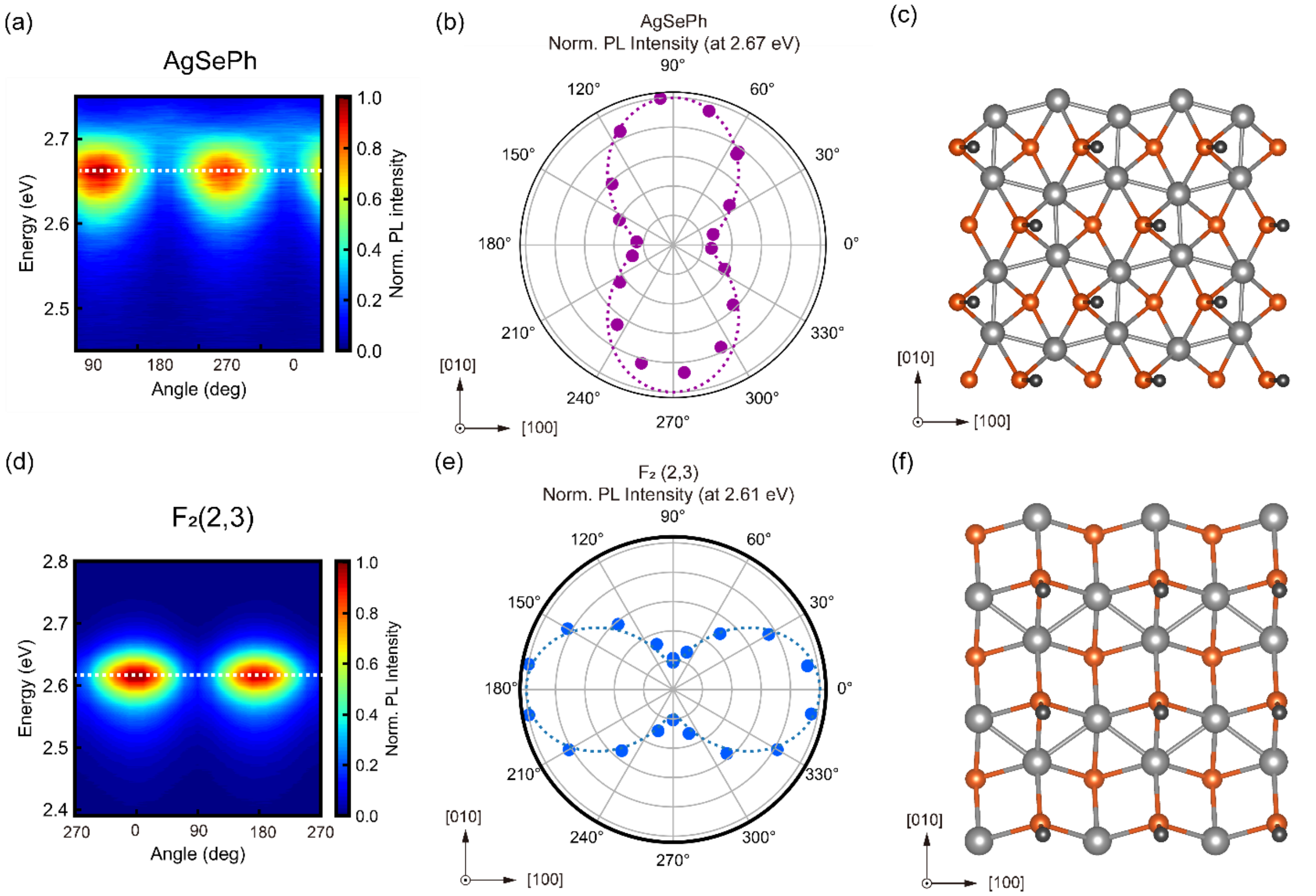

**Figure S13. Emission anisotropy of 2D MOCs.** (a) PL spectrum of **AgSePh** as a function of polarization angle of emitted light. (b) Polar plot of normalized PL intensity of **AgSePh** at 2.67 eV. (c) Structure of **AgSePh**. (d) PL spectrum of **$F_2$(2,3)** as a function of polarization angle of emitted light. (b) Polar plot of normalized PL intensity of **$F_2$(2,3)** at 2.61 eV. (c) Structure of **$F_2$(2,3)**.

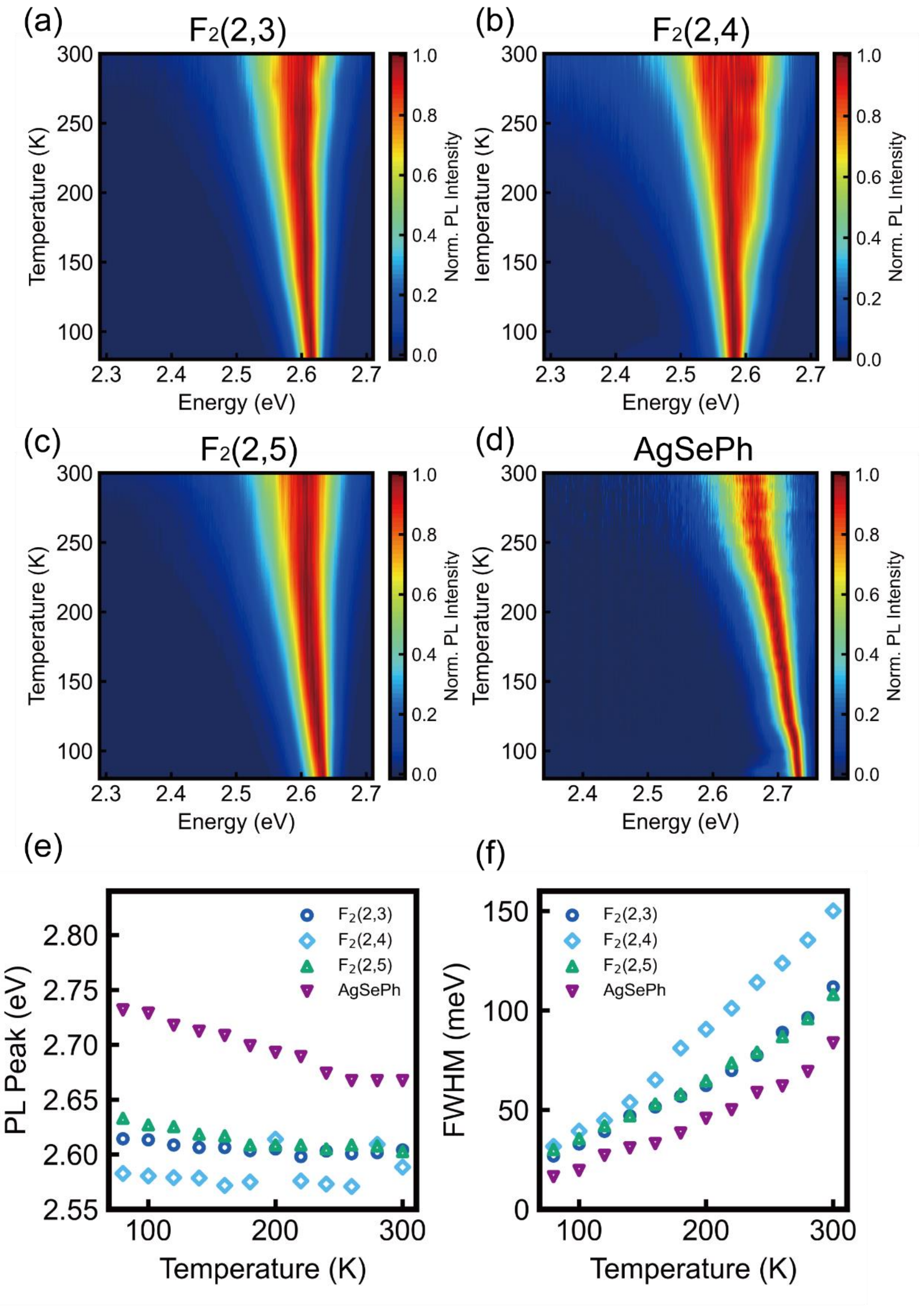

**Figure S14.** Normalized temperature dependent PL spectra of (a) **$F_2$(2,3)**, (b) **$F_2$(2,4)**, (c) **$F_2$(2,5)**, (d) **AgSePh** from 80 to 300 K. The data of AgSePh was reported in ref. 5. (e) Evolution of PL peak position and full-width at half-maxima as a function of temperature from 80 to 300 K.

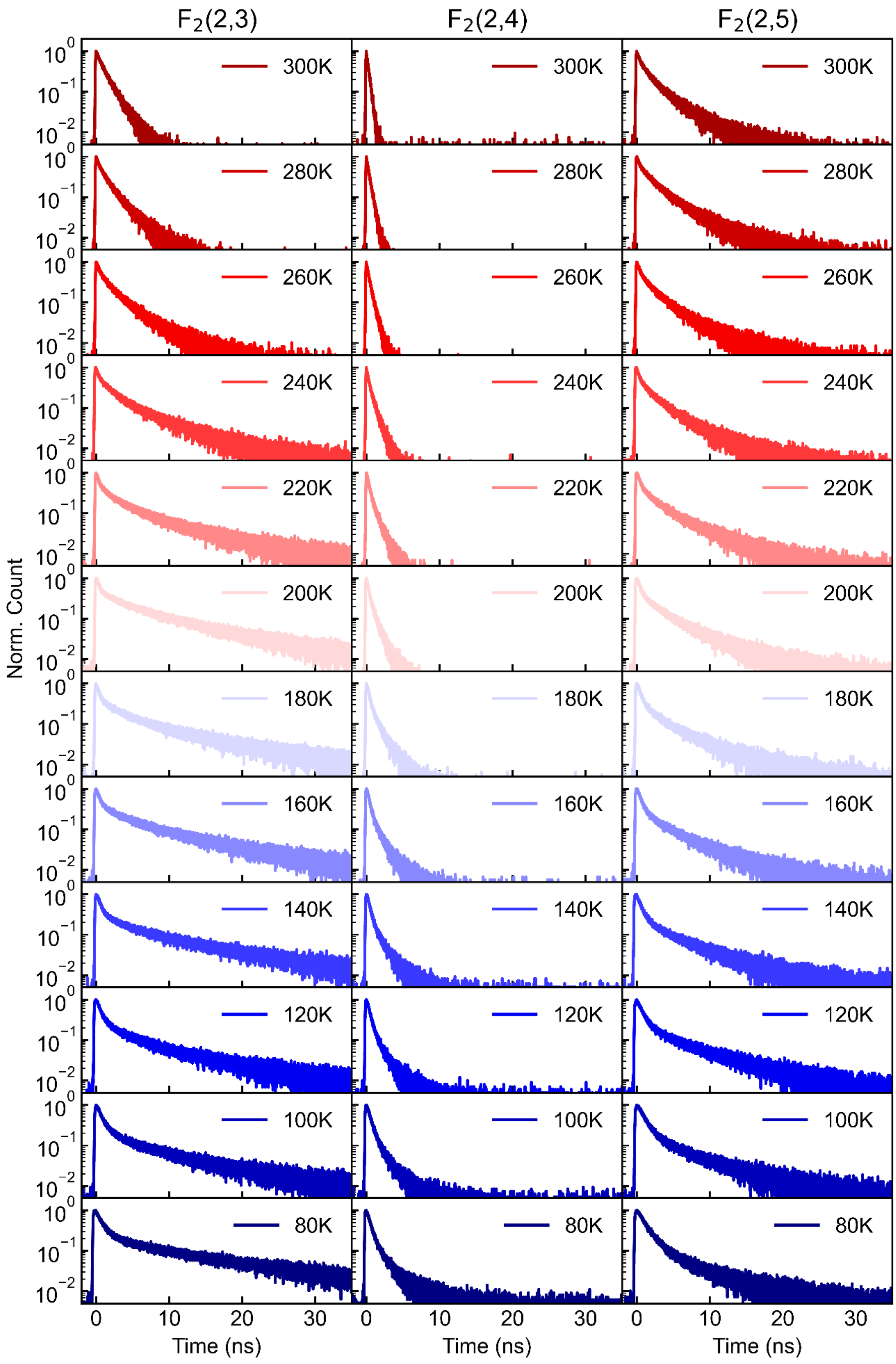

**Figure S15.** TRPL decay for **$F_2$(2,3)**, **$F_2$(2,4)**, and **$F_2$(2,5)** from 80 to 300 K.

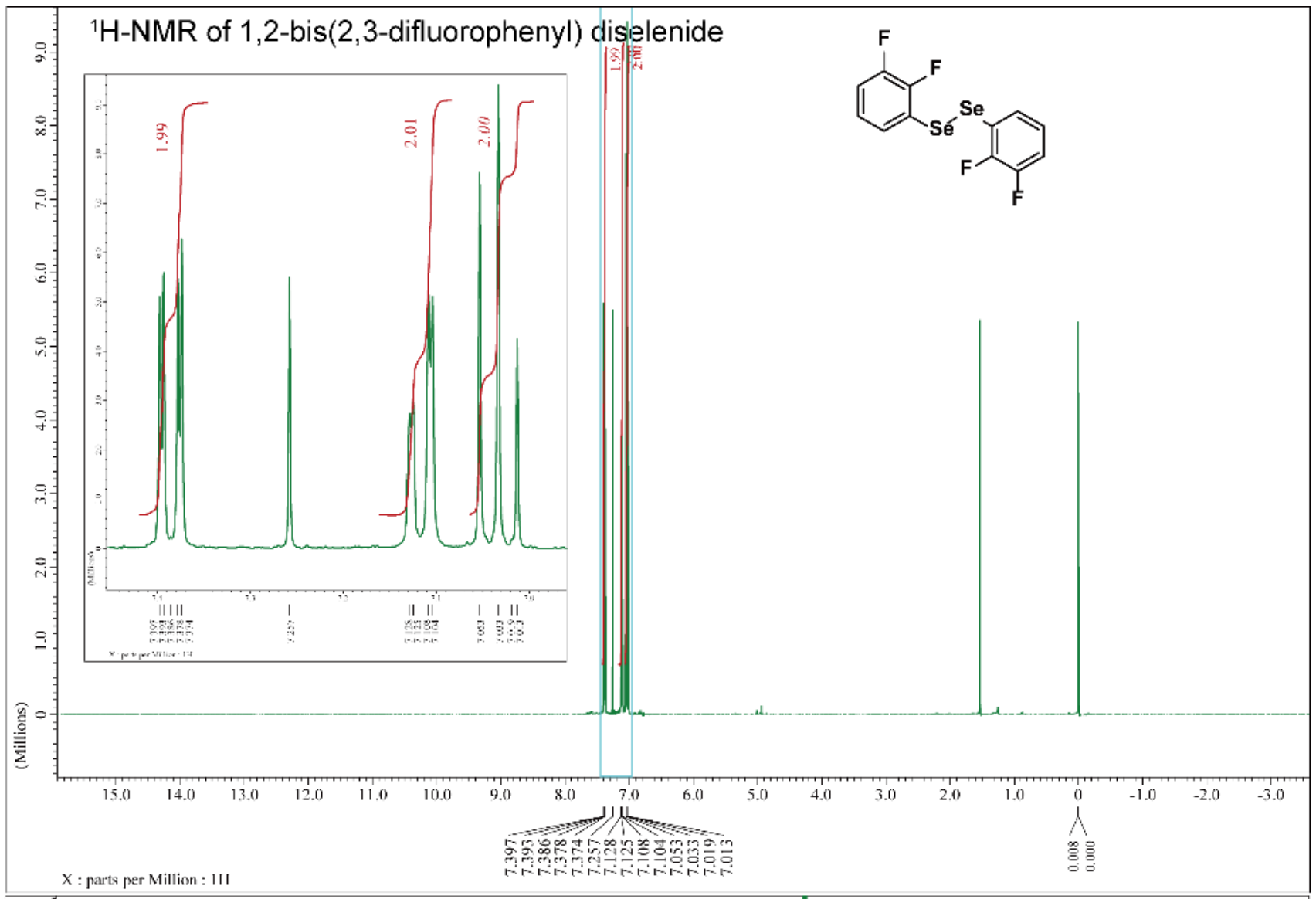

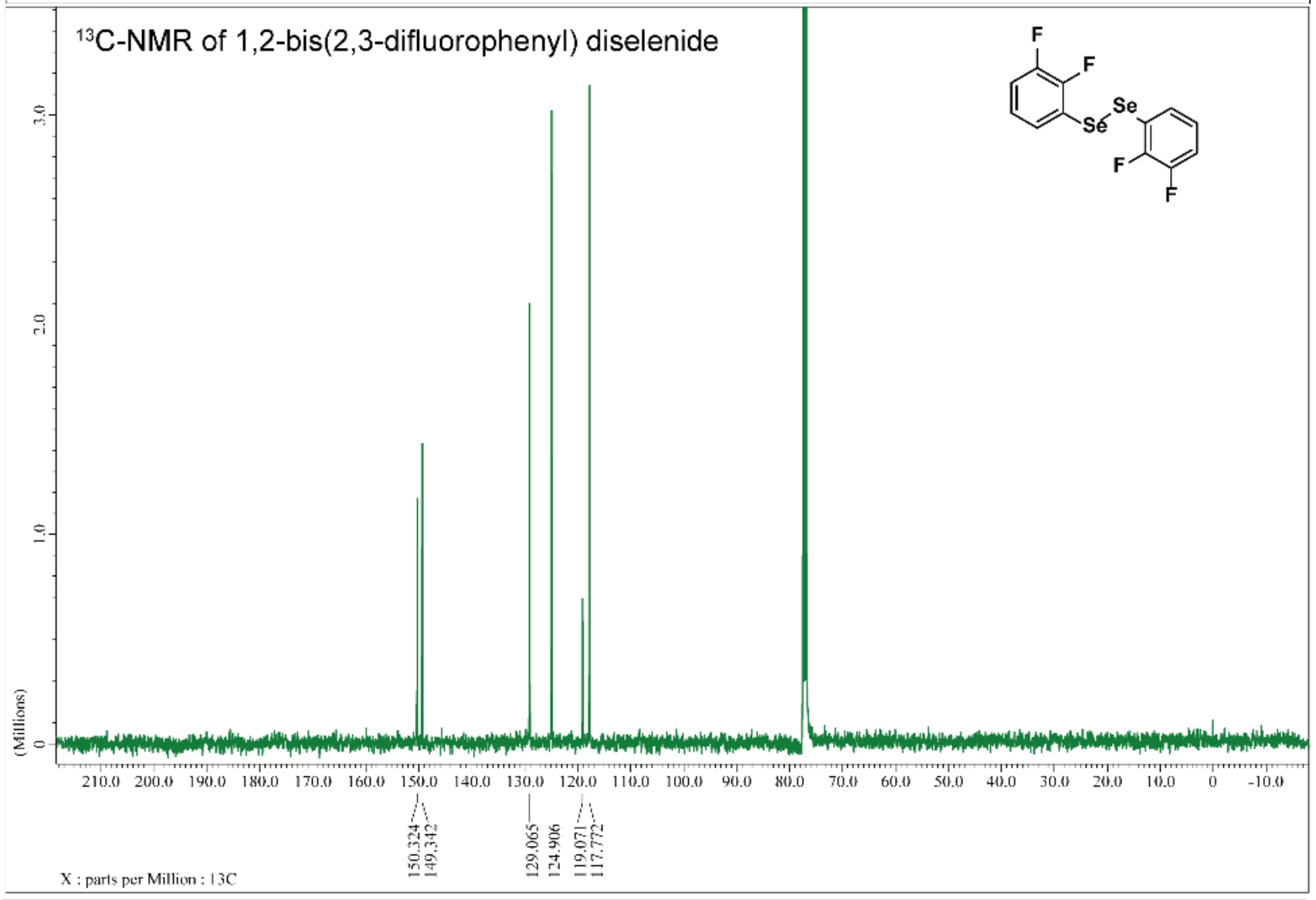

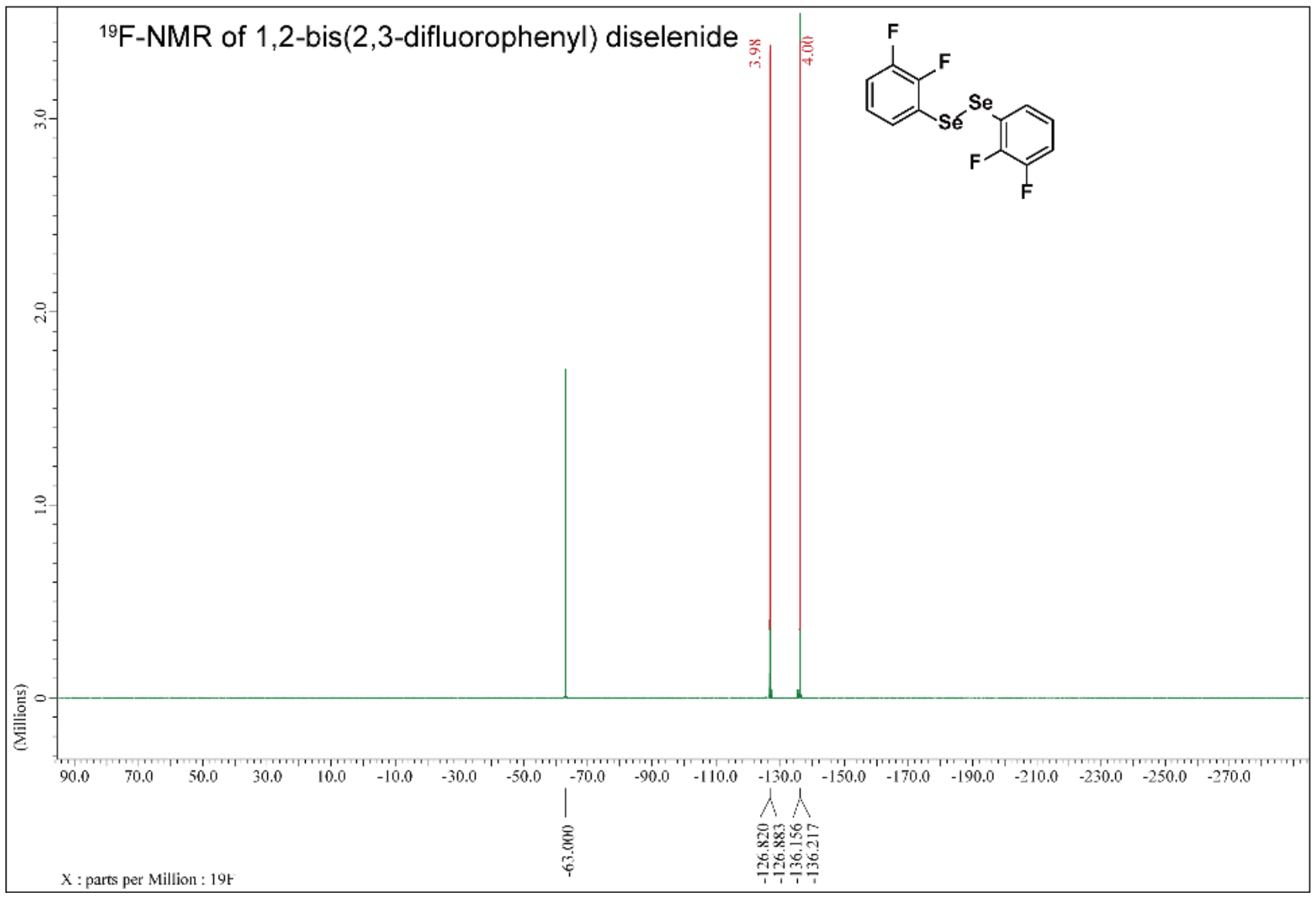

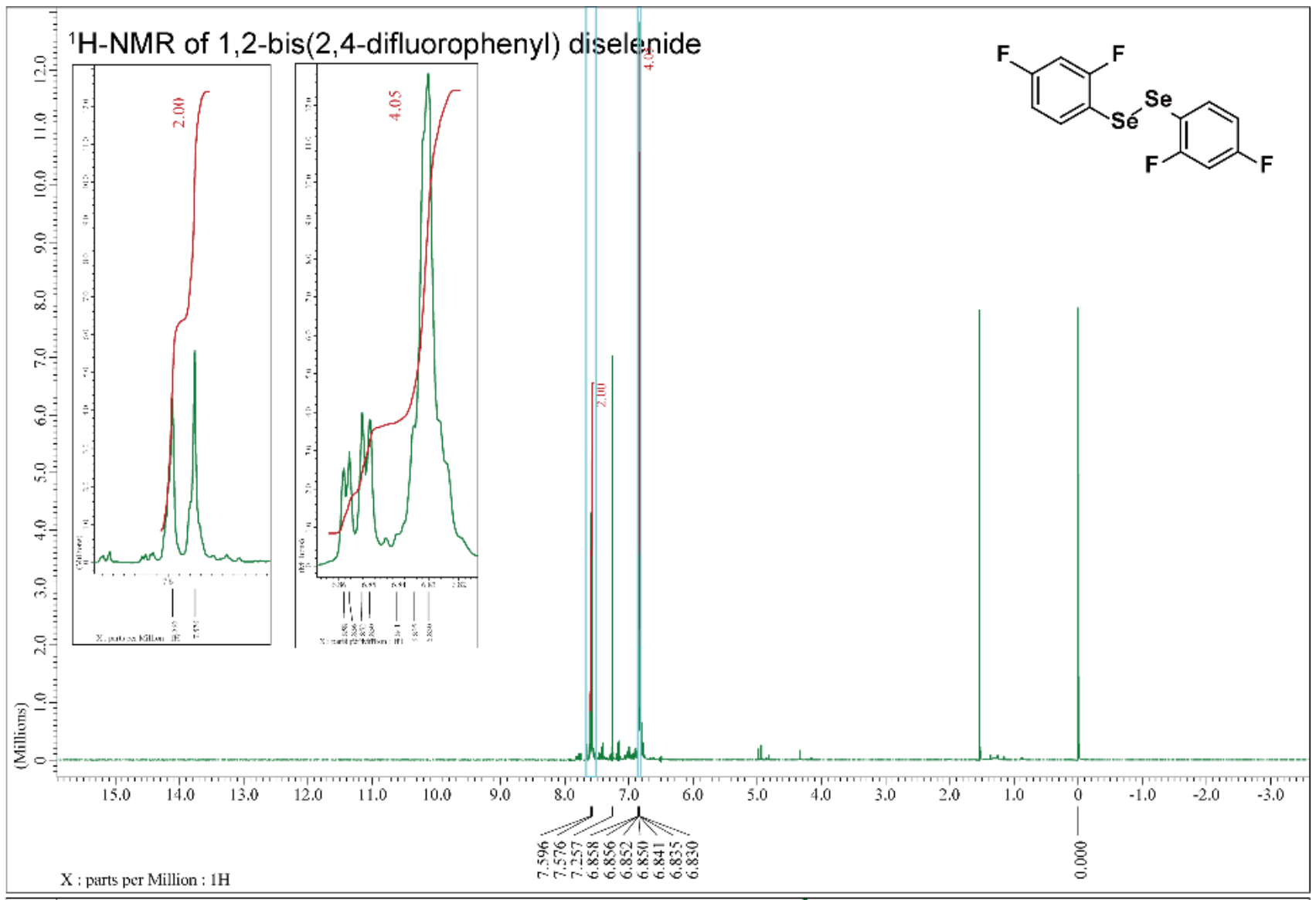

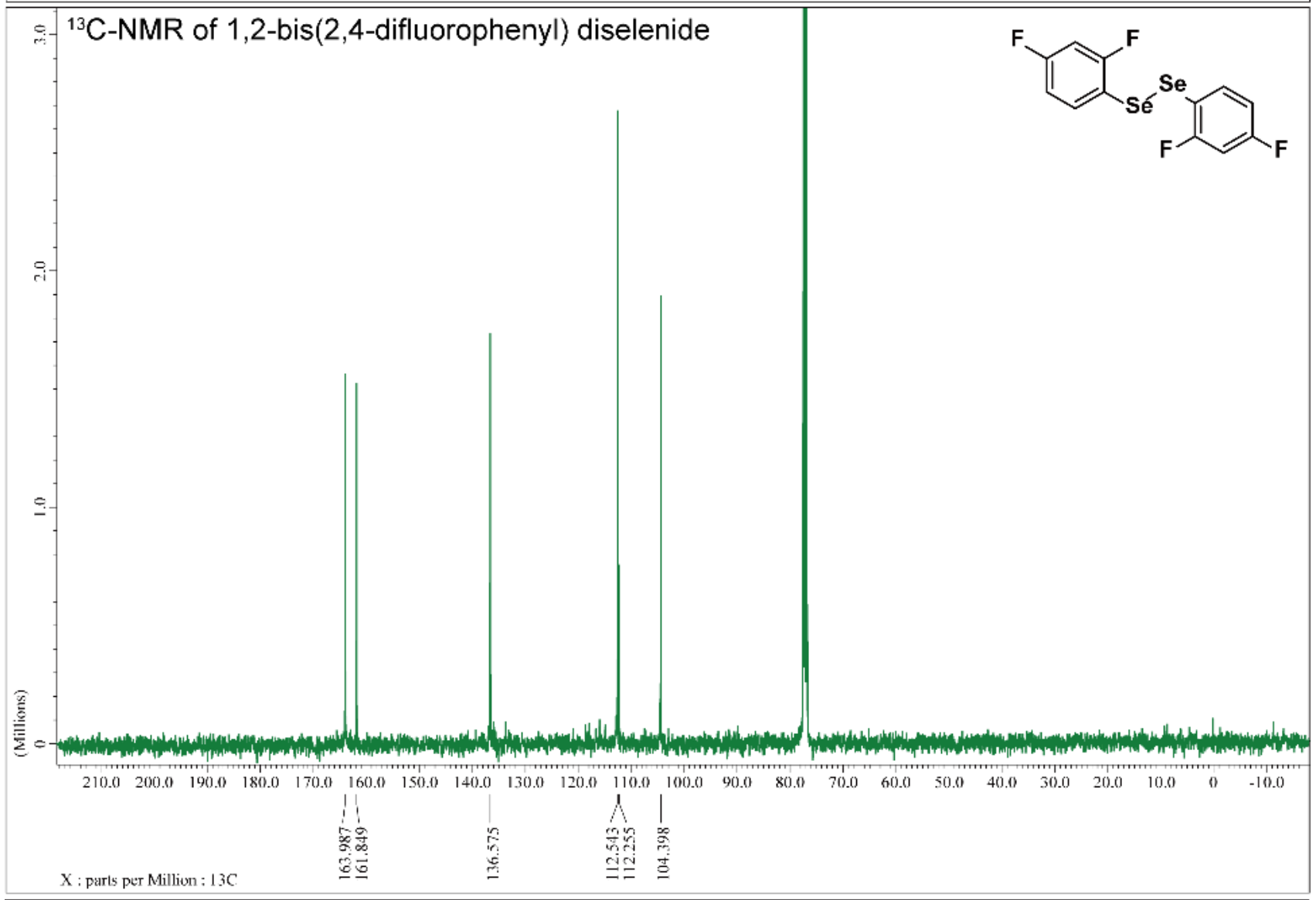

$^{13}$C-NMR of 1,2-bis(2,4-difluorophenyl) diselenide
X : parts per Million : 13C
163.987
161.849
136.575
112.543
112.255
104.398

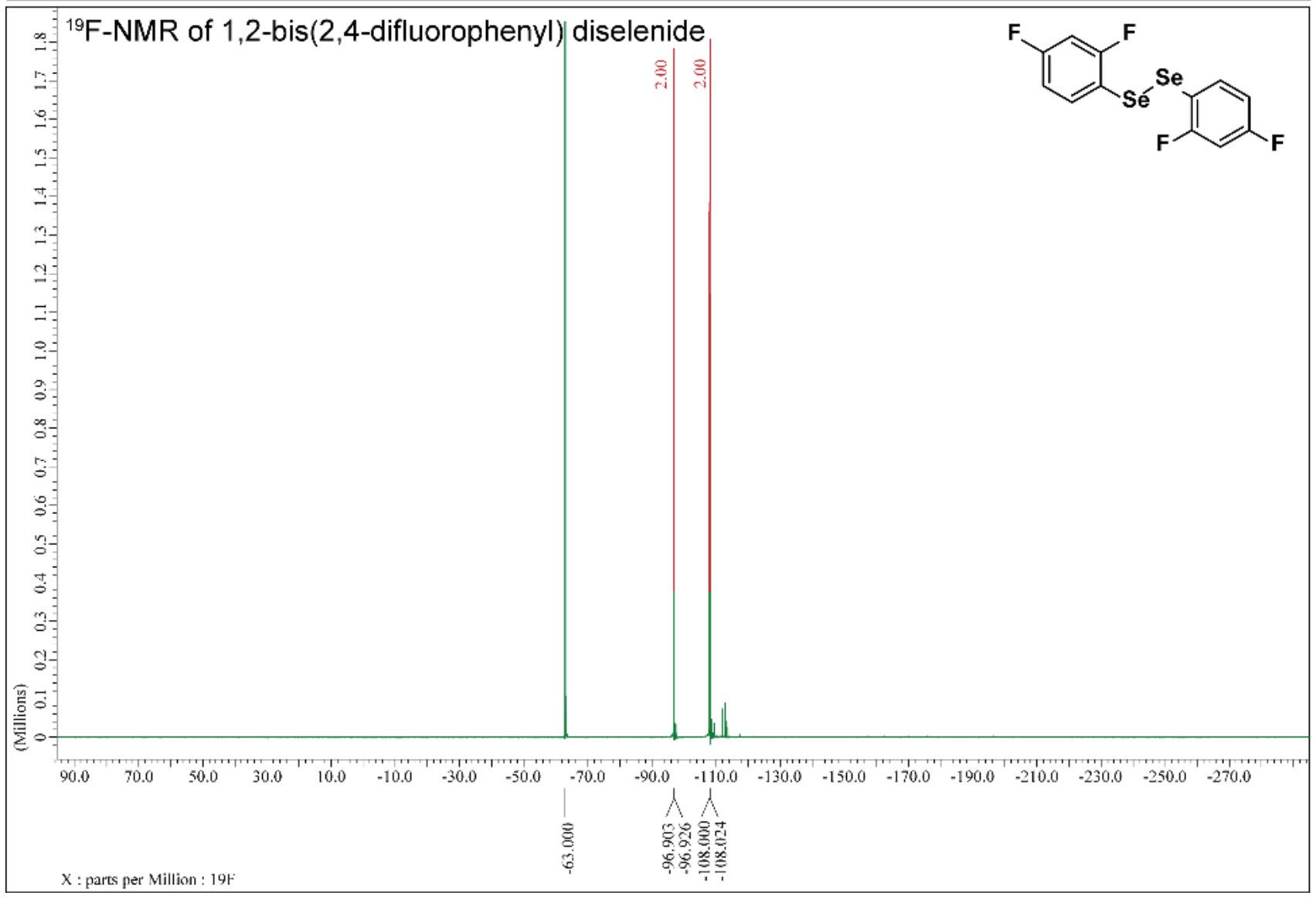

$^{19}$F-NMR of 1,2-bis(2,4-difluorophenyl) diselenide
X : parts per Million : 19F
-63.000
-96.903
-96.926
-108.000
-108.024

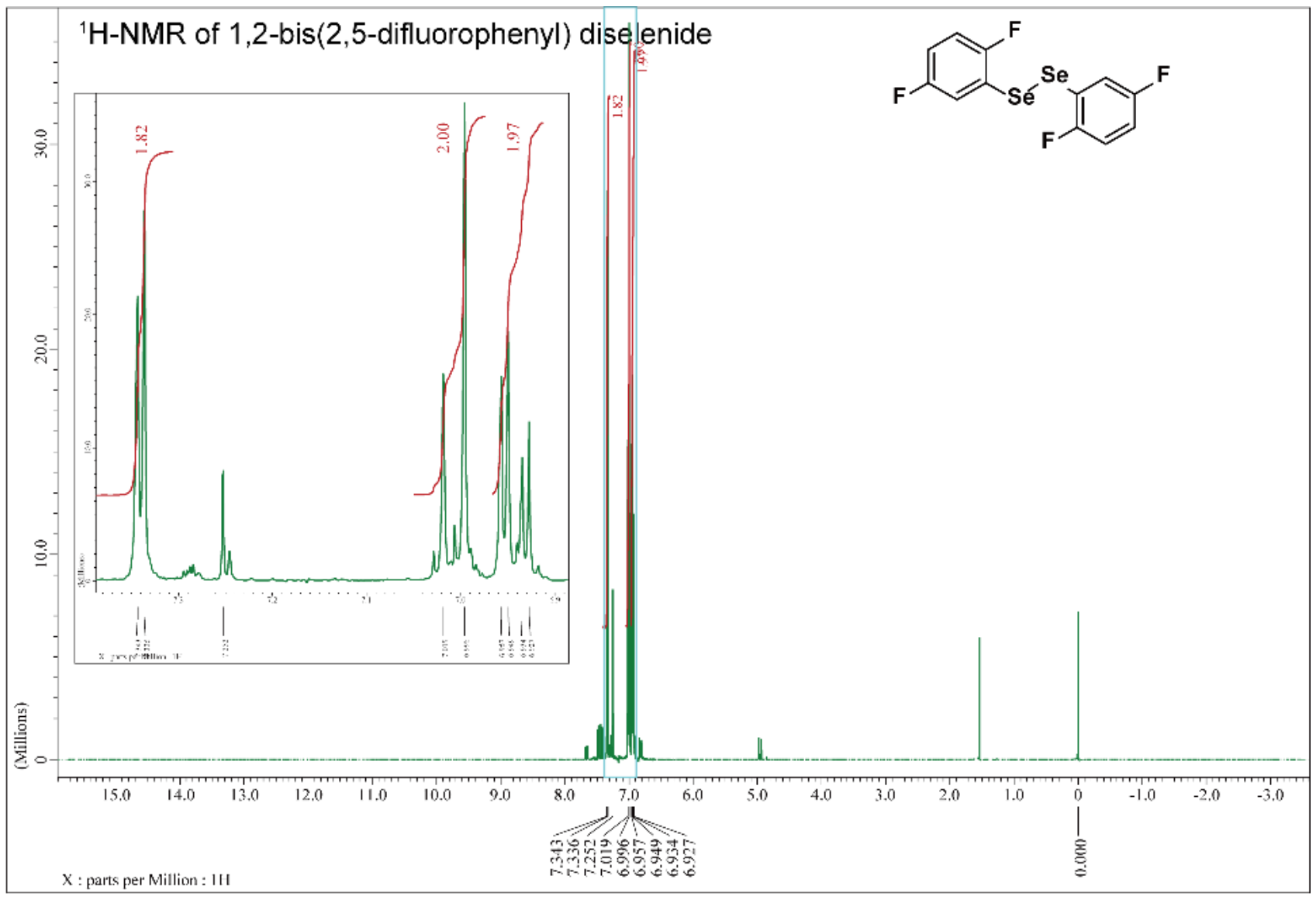

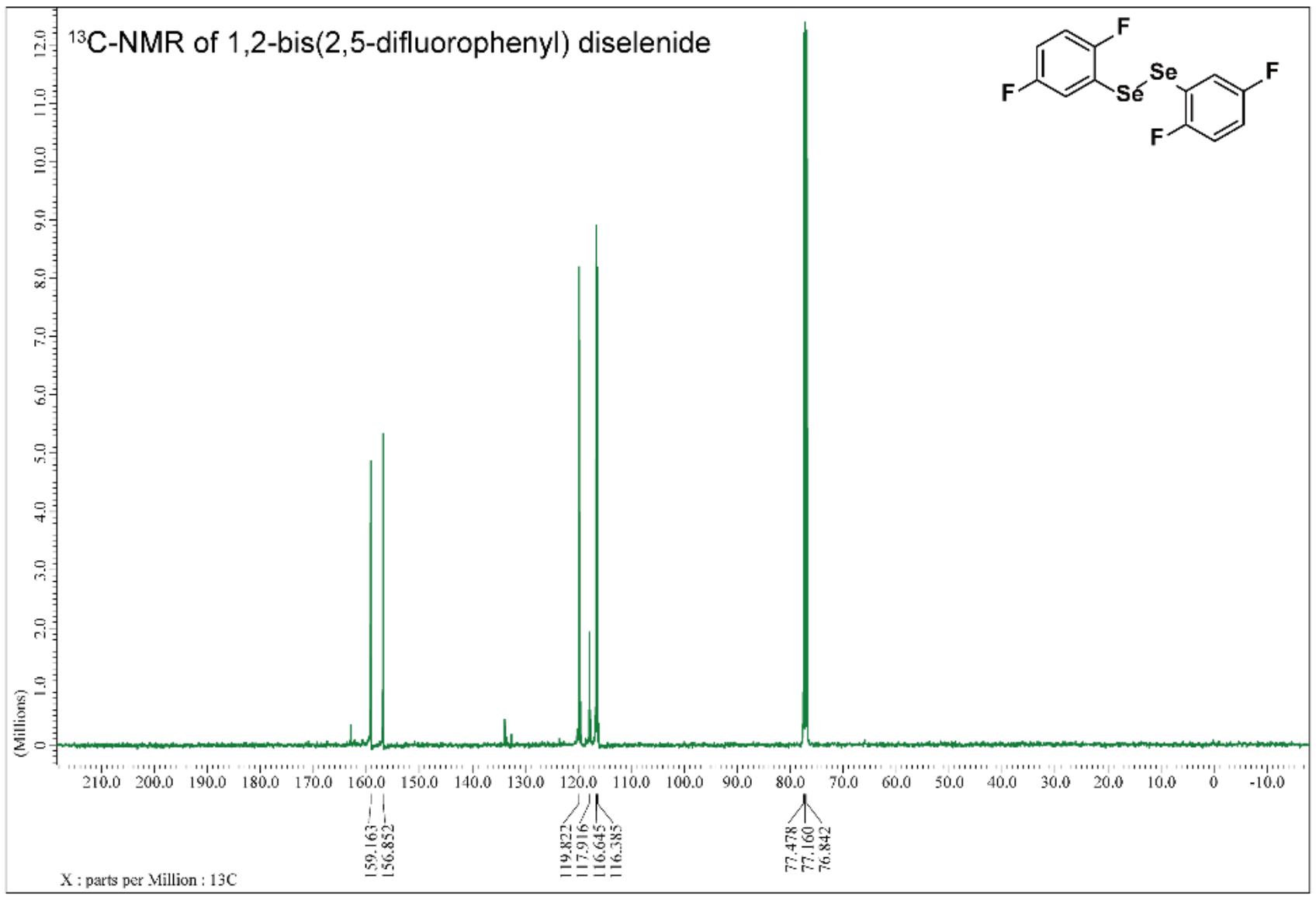

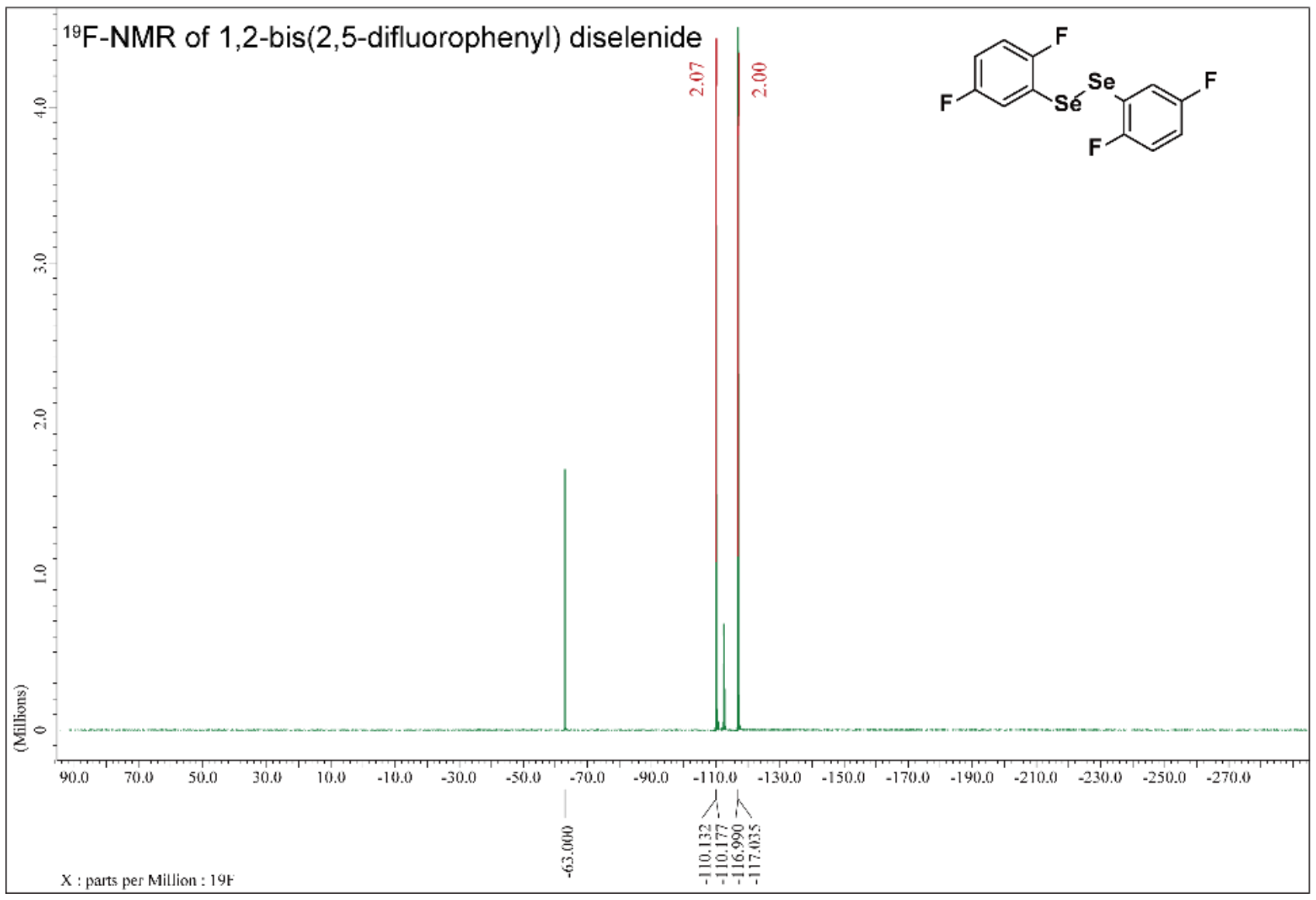
¹⁹F-NMR of 1,2-bis(2,5-difluorophenyl) diselenide
X : parts per Million : 19F
-63.000
-110.132
-110.177
-116.990
-117.035
2.07
2.00

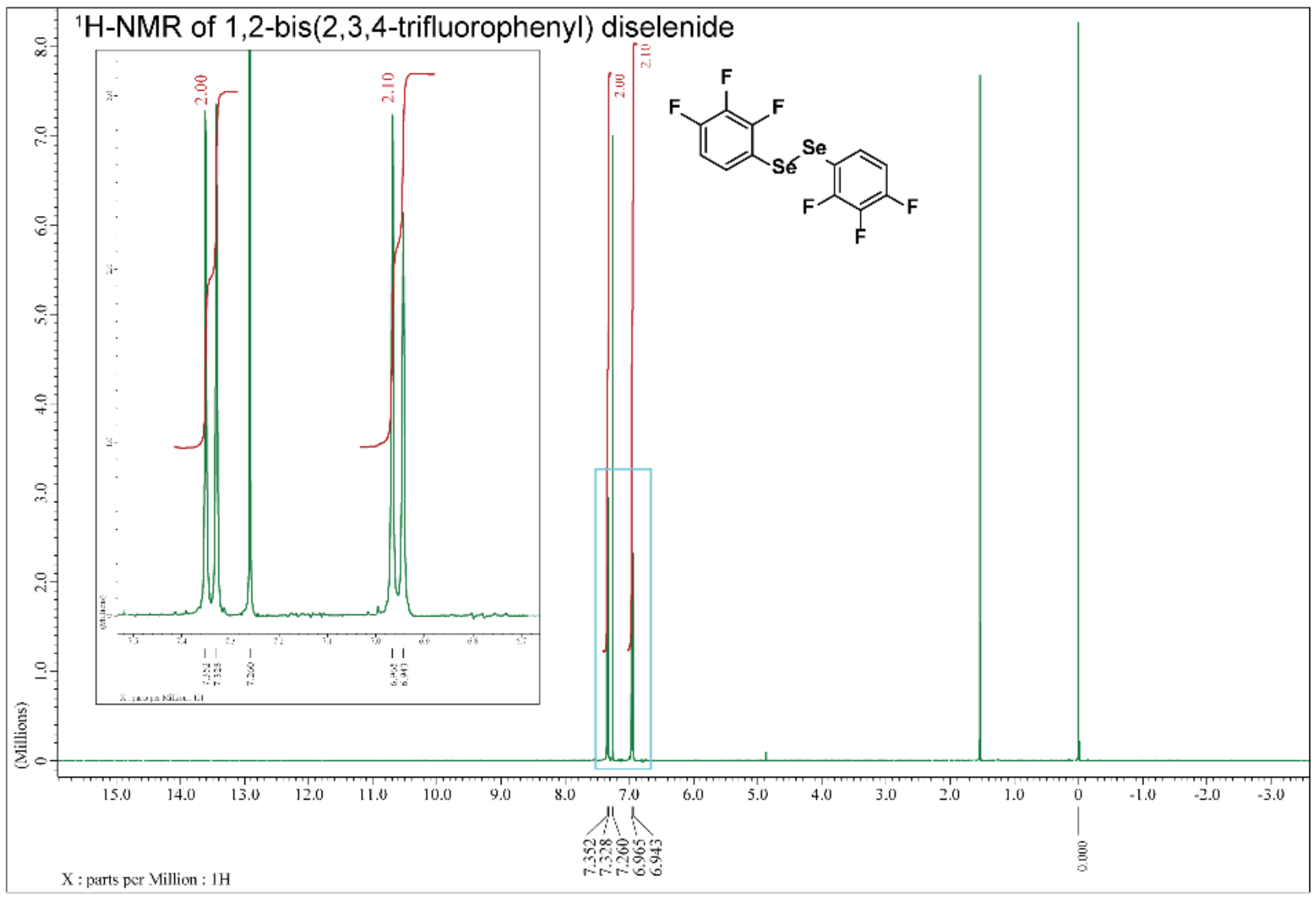

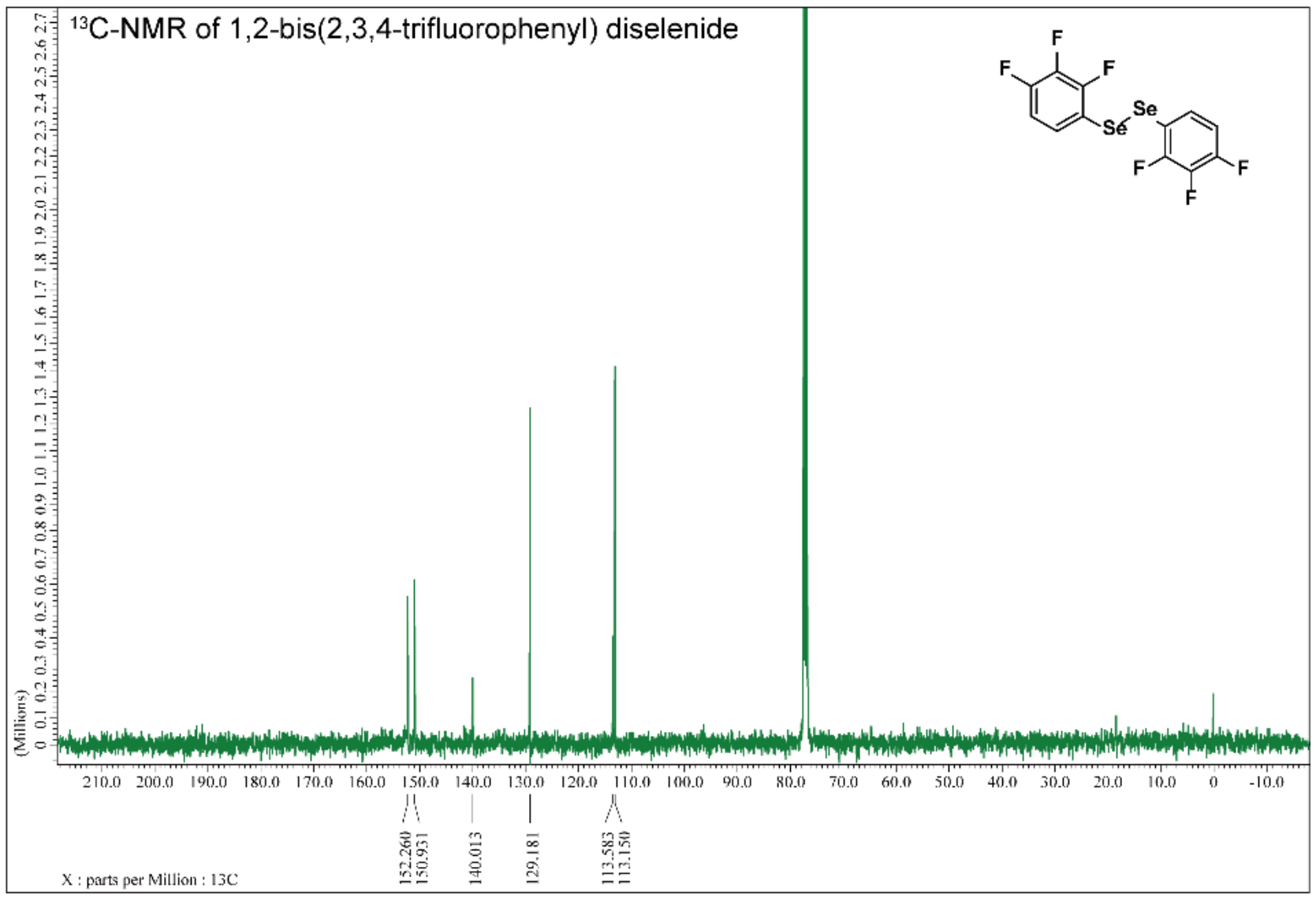

13C-NMR of 1,2-bis(2,3,4-trifluorophenyl) diselenide
152.260
150.931
140.013
129.181
113.583
113.150
X : parts per Million : 13C
(Millions)

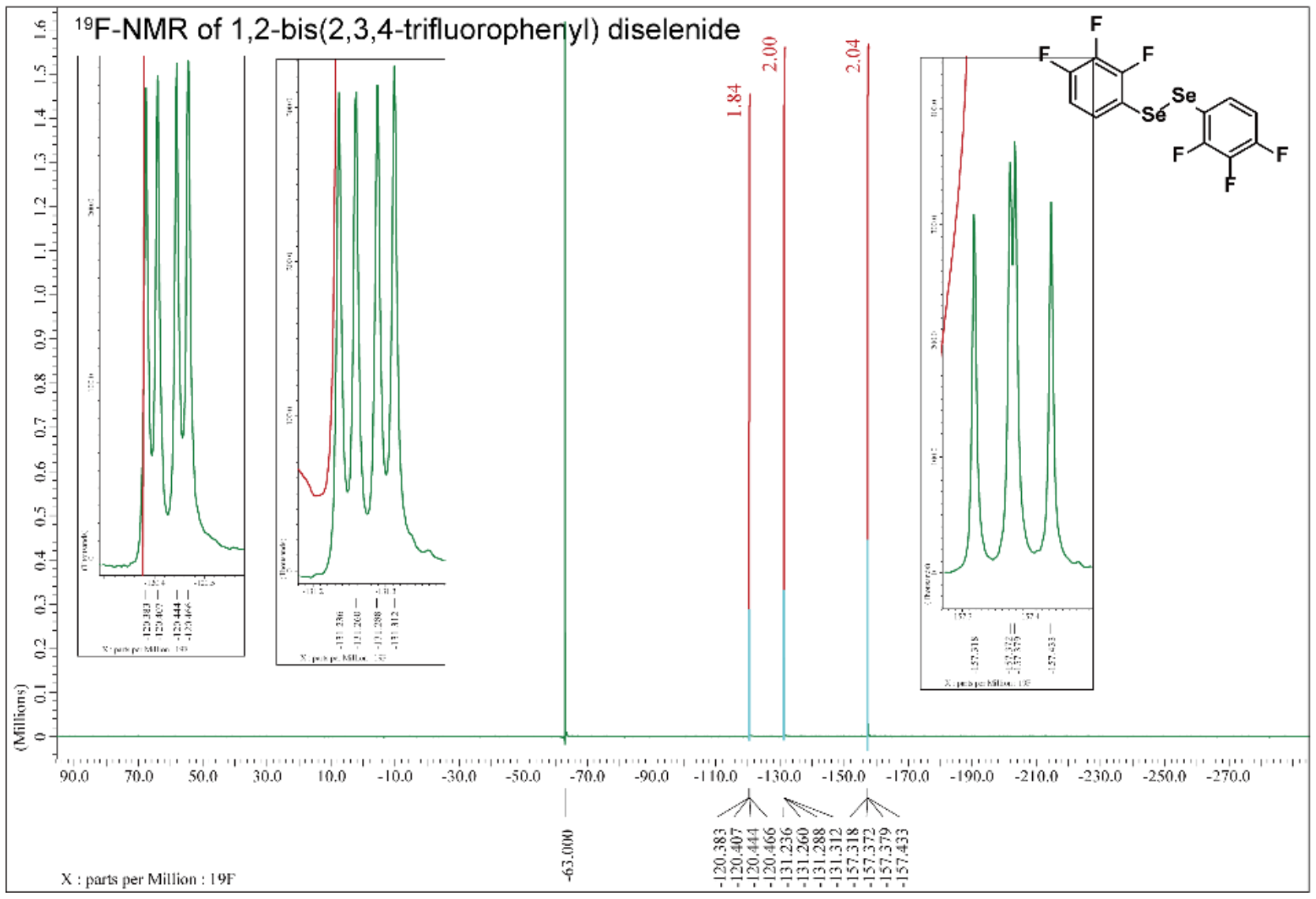

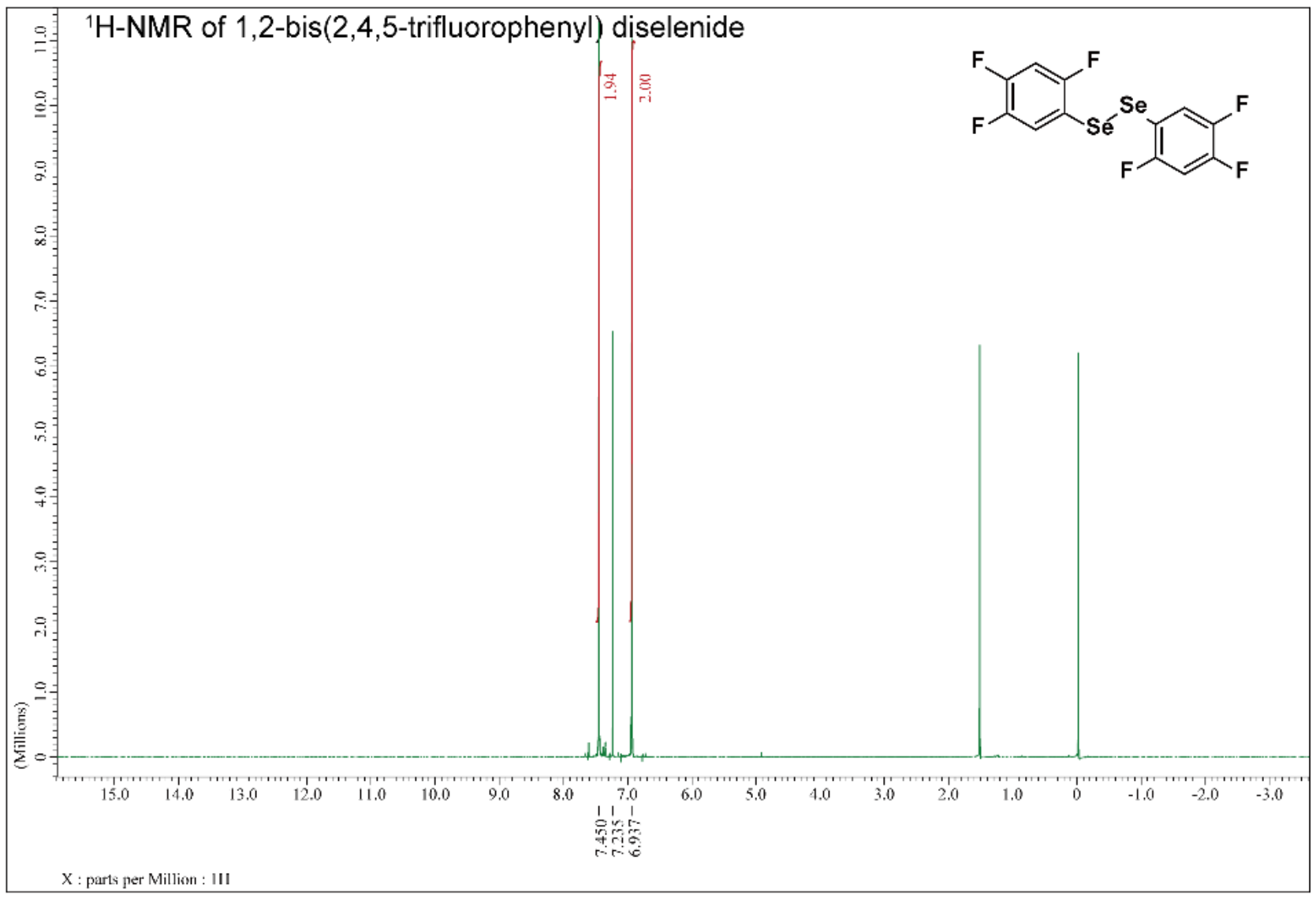
¹H-NMR of 1,2-bis(2,4,5-trifluorophenyl) diselenide
7.450
7.235
6.937
X : parts per Million : 1H

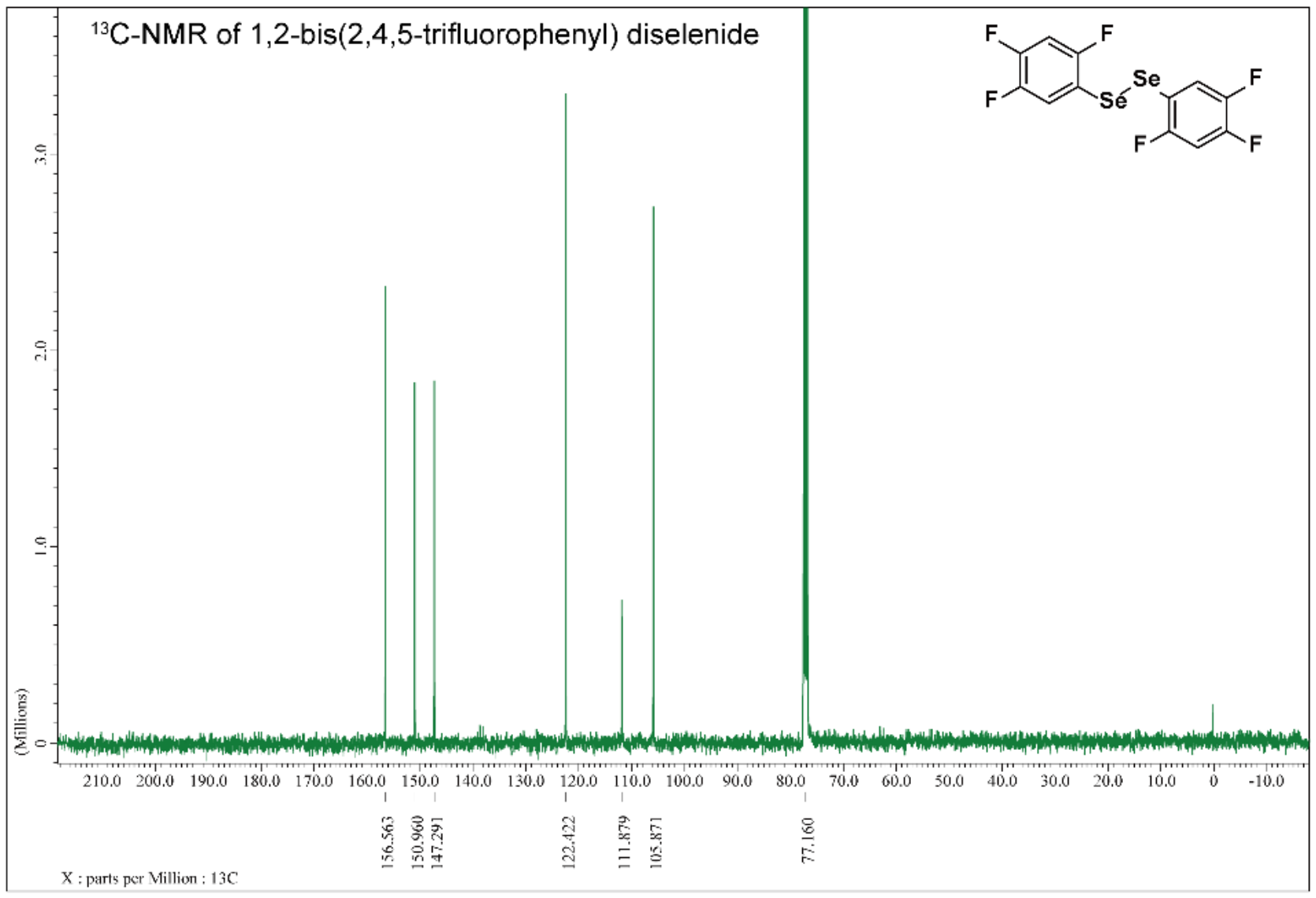
¹³C-NMR of 1,2-bis(2,4,5-trifluorophenyl) diselenide
156.563
150.960
147.291
122.422
111.879
105.871
77.160
X : parts per Million : 13C

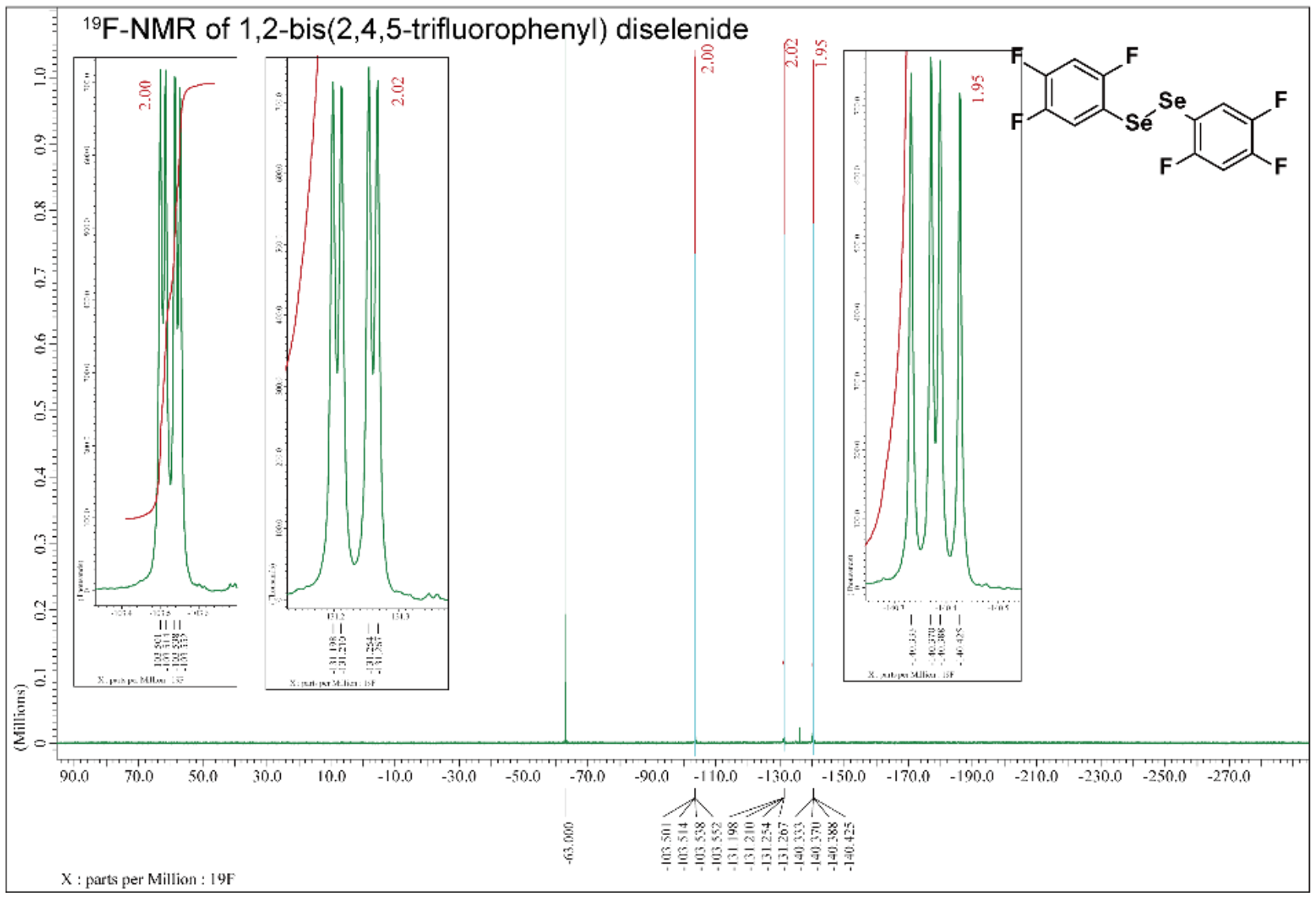

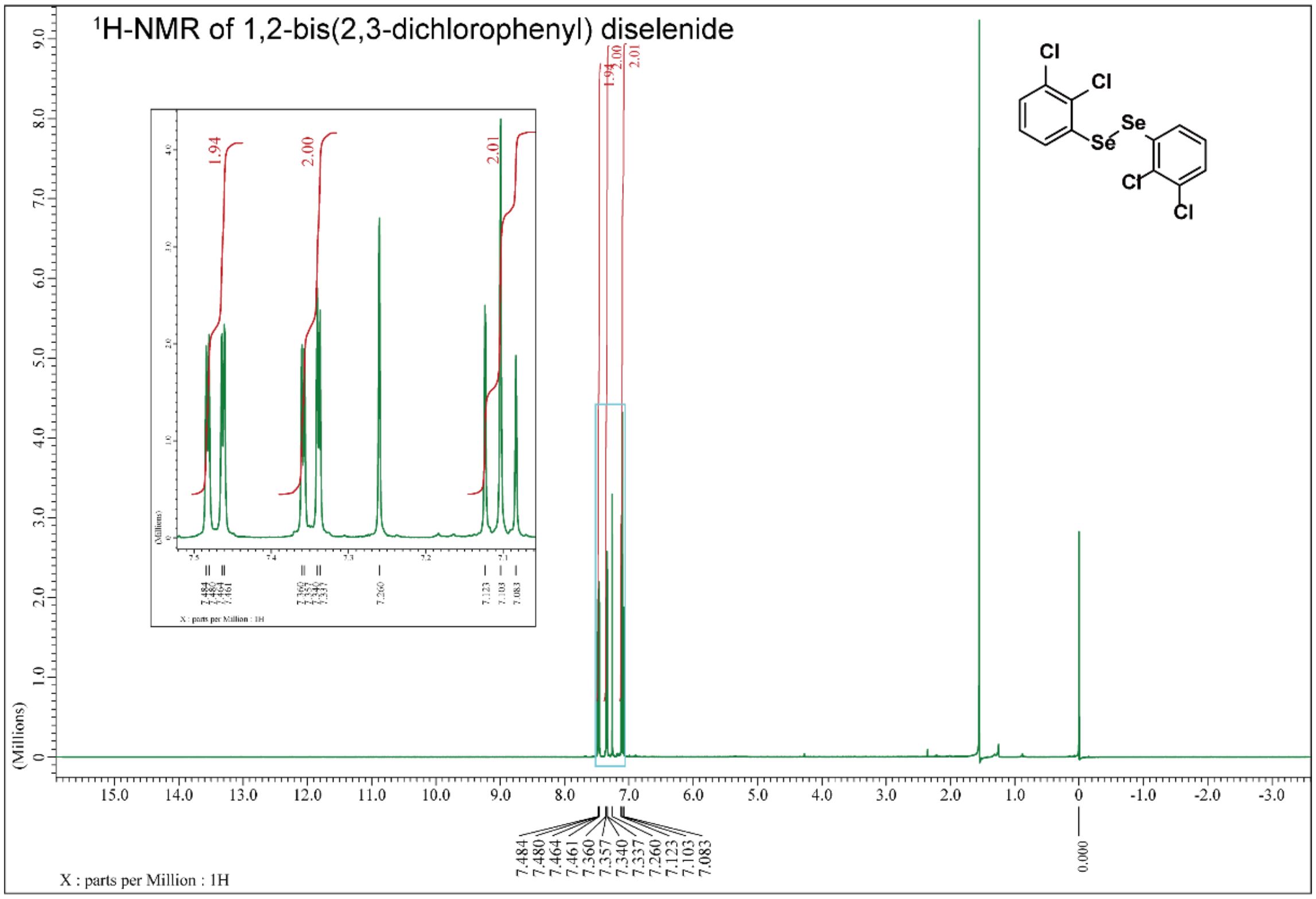

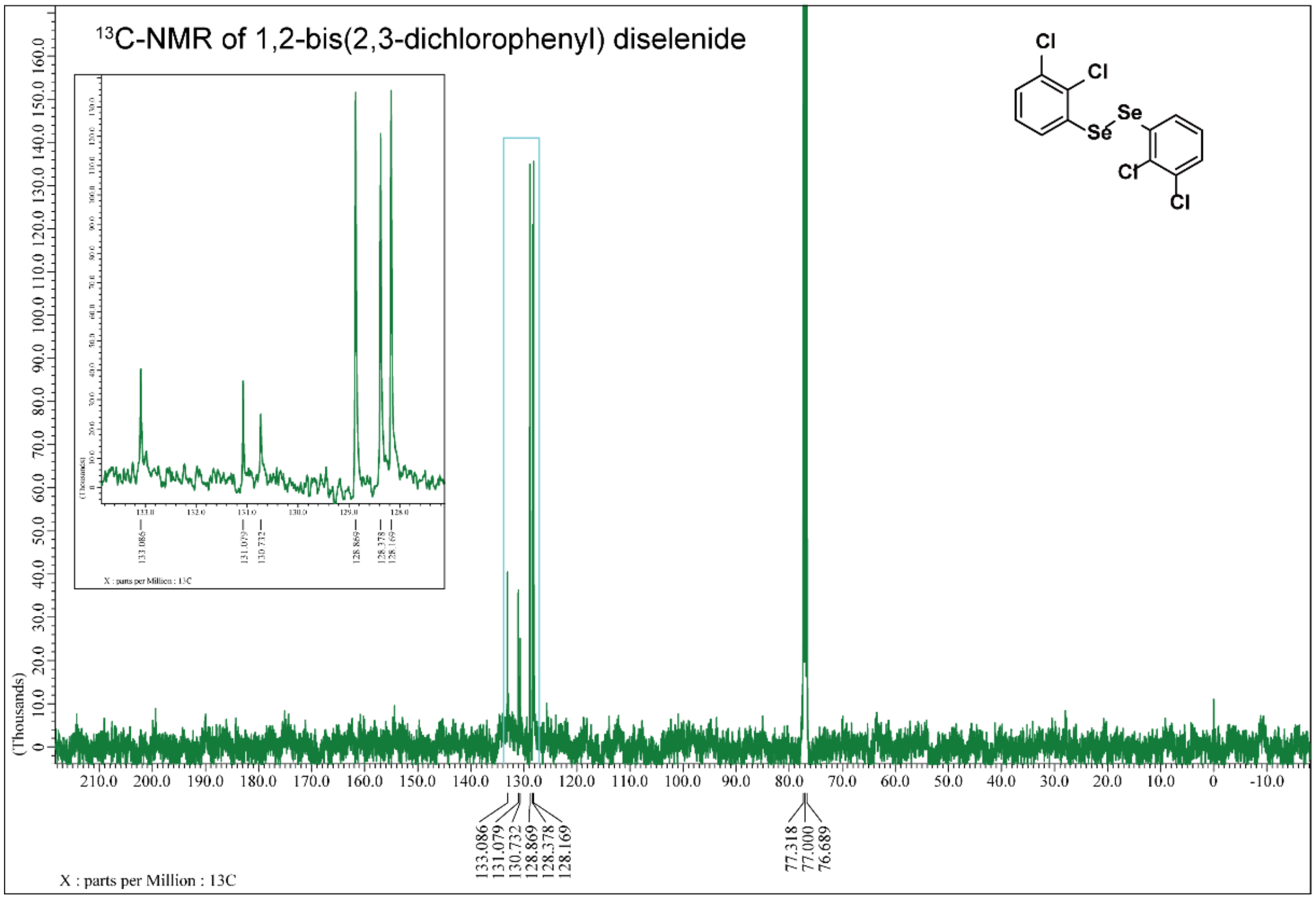